\documentclass[preprint,12pt]{elsarticle}

\usepackage{amssymb}
\usepackage{amsthm}
\usepackage{amsmath}
\usepackage{hyperref}
\usepackage{caption}
\usepackage{subcaption}
\usepackage{colortbl}
\usepackage{overpic}
\usepackage{diagbox}
\usepackage{booktabs}
\usepackage{array}
\usepackage{graphicx}
\usepackage{multirow}
\usepackage{booktabs}
\usepackage{textcomp}
\usepackage[title]{appendix}
\newcommand{\etal}{et al.}

\newcommand{\ph}[1]{{\bfseries \color{red} #1}}

\journal{Additive Manufacturing}

\begin{document}

\begin{frontmatter}



\title{PH-Net: Parallelepiped Microstructure Homogenization via 3D Convolutional Neural Networks}

%

\author[inst1]{Hao Peng}

\affiliation[inst1]{organization={School of Computer Science and Technology, Shandong University},
            addressline={72 Binhai Road}, 
            city={Qingdao},
            postcode={266237}, 
            state={Shandong Province},
            country={China}}
\affiliation[inst2]{organization={School of Mechanical Engineering, Shandong University},
            addressline={17923 Jingshi Road}, 
            city={Jinan},
            postcode={250061}, 
            state={Shandong Province},
            country={China}}

\author[inst1]{An Liu}
\author[inst1]{Jingcheng Huang}
\author[inst1]{Lingxin Cao}
\author[inst2]{Jikai Liu}
\author[inst1]{Lin Lu\corref{mycorrespondingauthor}}
\cortext[mycorrespondingauthor]{Corresponding author}
\ead{llu@sdu.edu.cn}


\begin{abstract}
Microstructures are attracting academic and industrial interests with the rapid development of additive manufacturing. The numerical homogenization method has been well studied for analyzing mechanical behaviors of microstructures; however, it is too time-consuming to be applied to online computing or applications requiring high-frequency calling, e.g., topology optimization.
Data-driven homogenization methods emerge as a more efficient choice but limit the microstructures into a cubic shape, which are infeasible to the periodic microstructures with a more general shape, e.g., parallelepiped. This paper introduces a fine-designed 3D convolutional neural network (CNN) for fast homogenization of parallel-shaped microstructures, named PH-Net.
Superior to existing data-driven methods, PH-Net predicts the local displacements of microstructures under specified macroscope strains instead of direct homogeneous material, motivating us to present a label-free loss function based on minimal potential energy. For dataset construction, we introduce a shape-material transformation and voxel-material tensor to encode microstructure type,base material and boundary shape together as the input of PH-Net, such that it is CNN-friendly and enhances PH-Net on generalization in terms of microstructure type, base material, and boundary shape. PH-Net predicts homogenized properties with hundreds of acceleration compared to the numerical homogenization method and even supports online computing. 
Moreover, it does not require a labeled dataset and thus is much faster than current deep learning methods in training processing. Benefiting from predicting local displacement, PH-Net provides both homogeneous material properties and microscopic mechanical properties, e.g., strain and stress distribution, yield strength, etc. We design a group of physical experiments and verify the prediction accuracy of PH-Net.
\end{abstract}

\end{frontmatter}


\section{Introduction}

Microstructures are ubiquitous in natural objects and possess a variety of excellent physical properties.
With the rapid development of additive manufacturing, microstructures are more and more utilized in industrial fields like mechanical, aerospace, and civil engineering, e.g., design for lightweight and high strength~\cite{lu2014build}.
Almost all materials possess heterogeneous and complex microstructures at a certain scale, contributing to superior mechanical, thermal, and electromagnetic properties.
However, understanding the behavior of such microstructures is not an easy task for their distinct geometry, volume fraction, and constituents properties. 
It is nearly impossible to experiment on a large number of microstructural samples with different geometry and physical properties, attributed to time and cost consumption. Also, simulating the entire body leads to computationally expensive and high memory storage requirements~\cite{saeb2016aspects}.

To address these problems, homogenization theory has been developed as a multiscale technique in past decades.
The homogenization theory, originally developed to solve partial differential equation (PDE) problems, is to estimate the homogenized macroscopic properties from the response of its underlying microstructure, thereby allowing to substitute the heterogeneous material with an equivalent homogeneous one~\cite{saeb2016aspects,rao2020three}.
These approaches have three main categories. 
(i) Analytical methods, i.e., Voigt and Reuss assumptions~\cite{yoigt1889uber,reuss1929berechnung} specified the upper and lower bound of the macroscope properties.
While universal and very simple, they only provide rough estimates of the overall material properties and are unreliable for complex nonlinear structures.
(ii) Semi-analytical methods, e.g., Mori–Tanaka model~\cite{mori1973average,benveniste1987new}, the self-consistent scheme~\cite{kroner1958berechnung,willis1977bounds}, the generalized self-consistent scheme~\cite{kerner1956elastic,chatzigeorgiou2012effective}, and the differential method~\cite{mclaughlin1977study,norris1985differential}, which are mainly based on the mean-field approximation and approximate the interaction between the phases.
(iii) Numerical methods have been introduced to perform various analyses over the representative volume element (RVE), such as Voronoi cell finite element~\cite{ghosh1995elastic,moorthy1996model}, fast Fourier transform (FFT)~\cite{moulinec1998numerical}, boundary element (BE) method~\cite{kaminski1999boundary}, and finite element (FE) discretization~\cite{renard1987etude,moes2003computational,bouhala2014advanced}.
The boundary conditions in numerical homogenization are commonly considered with periodic displacement and anti-periodic traction. 
Although the numerical method is known to be computationally expensive, it has been more focused since it has shown to be adequate for high heterogeneous materials and complex microstructures in a high prediction accuracy.

The numerical homogenization methods have been developed as a basis of multiscale design applications, i.e., topology optimization~\cite{huang2011topological,zhu2017two,gao2018topological,zhang2021novel}, and functional structural design~\cite{meza2014strong,schumacher2015microstructures,panetta2015elastic,Martinez2016}, in which the input object is first partitioned into a voxel grid as a coarse scope, then the periodic microstructures are filled into the grid with homogeneous material properties.
However, such high-frequency calling of numerical homogenization method makes it highly time-consuming for simulation and optimization in those applications.
Hence, many works have been recently developed to decrease the computational cost of the multiscale analysis, which can be roughly classified into explicit and implicit methods.
Explicit methods~\cite{schumacher2015microstructures,panetta2015elastic,Martinez2016,temizer2007adaptive,yvonnet2009numerically,yvonnet2013computational,liu2021mechanical} build an explicit microstructure-to-material space by numerical homogenization method offline, then query target homogenized material properties via interpolation.
Limited by interpolation method, these methods require a mass of dense and mesh-grid-like samples to ensure accuracy. 
Plus, these methods do not support interpolation between different microstructures, which means each explicit space is only applicable to a specific type of microstructures.
Machine learning (ML) and deep learning (DL) methods have been introduced to microstructure analysis. 
The main idea of these methods is to build an implicit homogenization predictor instead of using explicit interpolation.
Many works \cite{fritzen2018two,lookman2019active,ford2021machine} based on ML methods have successful applications in terms of homogenization prediction, navigation, and two-scale structure modeling.
A multi-type of microstructure can be embedded in the same implicit space in these methods, but heavily relying on feature engineering and expert knowledge makes their applications limited.
Recently, DL approaches showed significant success in many applications due to the capability to search for the most salient features to be learned automatically. Hence, many research attempts utilized DL methods, such as neural networks (NN)~\cite{le2015computational,lu2019data}, convolutional neural networks (CNN)~\cite{yang2019establishing,rao2020three}, and graph convolutional networks (GCN)~\cite{vlassis2020geometric}, to predict the homogenized material properties.

Although current ML and DL methods generalize homogenization prediction with different types of microstructure, they still have some limitations, i.e., all presented data-driven homogenization methods still require numerical homogenization method to build ground-truth labels, which is the most time-consuming part. 
Plus, only voxel-based microstructures are implemented as the input of current data-driven methods, and the base material of microstructures only depends on ground-truth labels. 
It means the ground-truth labels, as well as corresponding homogenization predictor, have to be calculated and trained again if we change the base material of microstructures, lacking the generalization to different base materials, e.g., plastic, resin, metal, and so on.
Besides, with voxel-based microstructures as input, above data-driven methods cannot predict the homogeneous materials of microstructures with general but periodic boundary shape, e.g., parallelepiped. 
This makes above data-driven approaches to a narrow application of voxel-base multiscale microstructure modeling problems, which faces the inherent drawbacks that the clipped boundary voxels induce fragment structures and significant simulation errors compared to hex-based multiscale frameworks.

Recently, Tozonoi~\etal~\cite{low-parametric} introduced a rhombic microstructure family in a 2D plane. They built an explicit geometry-material space for a two-scale framework, in which both microstructure parameters and a shape parameter (rhombic angle $\alpha$) were used for structural design.
However, this method is somewhat limited as it only suits isotropic microstructures.
Moreover, it is extreme timing-cost to extent to parallelepiped microstructures directly with the time increase on both 3D homogenization computation and dense microstructure sampling on extra shape parameters ($1\to6$).

We introduce a 3D convolutional neural network called PH-Net to predict the homogeneous material properties with high-efficiency for arbitrary periodic microstructures with a parallel hexahedral boundary shape. 
Unlike the current data-driven homogenization method, we regard PH-Net as an implicit PDE solver to output the microscopic displacement fields of microstructures under specified macroscopic strain field and present a label-free loss function based on minimum potential energy (MPE) theory. Compared to existing ML/DL methods that use mean squared (MSE) loss function, the dataset of PH-Net is allowed to be generated without any ground-truth, since computing ground-truth labels using the numerical homogenization method could be the most time-consuming part for microstructures with parallel hexahedral boundary shape.
Instead of inputting a binary voxel matrix as the above methods, we construct a material-voxel tensor, in which both base material and microstructure information are embedded together, as the input of PH-Net and perform a shape-material transformation to encode the boundary shape parameters of a microstructure into its base material.
Thus PH-Net shows superiority over other data-driven methods in terms of generalization of base materials and boundary shapes.
PH-Net speeds up hundreds of times than numerical homogenization methods and even achieves real-time predicting.
Benefiting from the prediction of displacements, PH-Net can provide more mechanical properties (e.g., strain and stress distribution, yield strength, and shear strength) than existing DL methods, which only can predict homogenized material properties.
Furthermore, our approach has better generalization performance, e.g., PH-Net performs better to predict homogenized material properties for such microstructures not involved in the dataset. 

The main contributions of our work are as follows.
\begin{itemize}
    \item We propose a 3D convolutional neural network based on 3D U-Net, named PH-Net, and a novel loss function based on minimum potential energy theory to predict the homogenized material properties of microstructures in a parallelepiped shape. PH-Net is label-free, more time-efficient than current deep learning methods, and of high generalization to arbitrary microstructures.
    \item We introduce a highly generalized dataset construction method, taking microstructure types and their base material and boundary shape into account via voxel-material tensor and shape-material transformation, which is backwards compatible with existing DL methods.
    \item  We predict the displacements for parallelepiped microstructures and more microscope mechanical properties, e.g., strain and stress distribution, yielding strength, and shear strength, besides homogenized material properties.
\end{itemize}

\section{Methodology}

\begin{figure*}
    \centering
     \centering
    \begin{overpic}[width=\linewidth,tics=10]{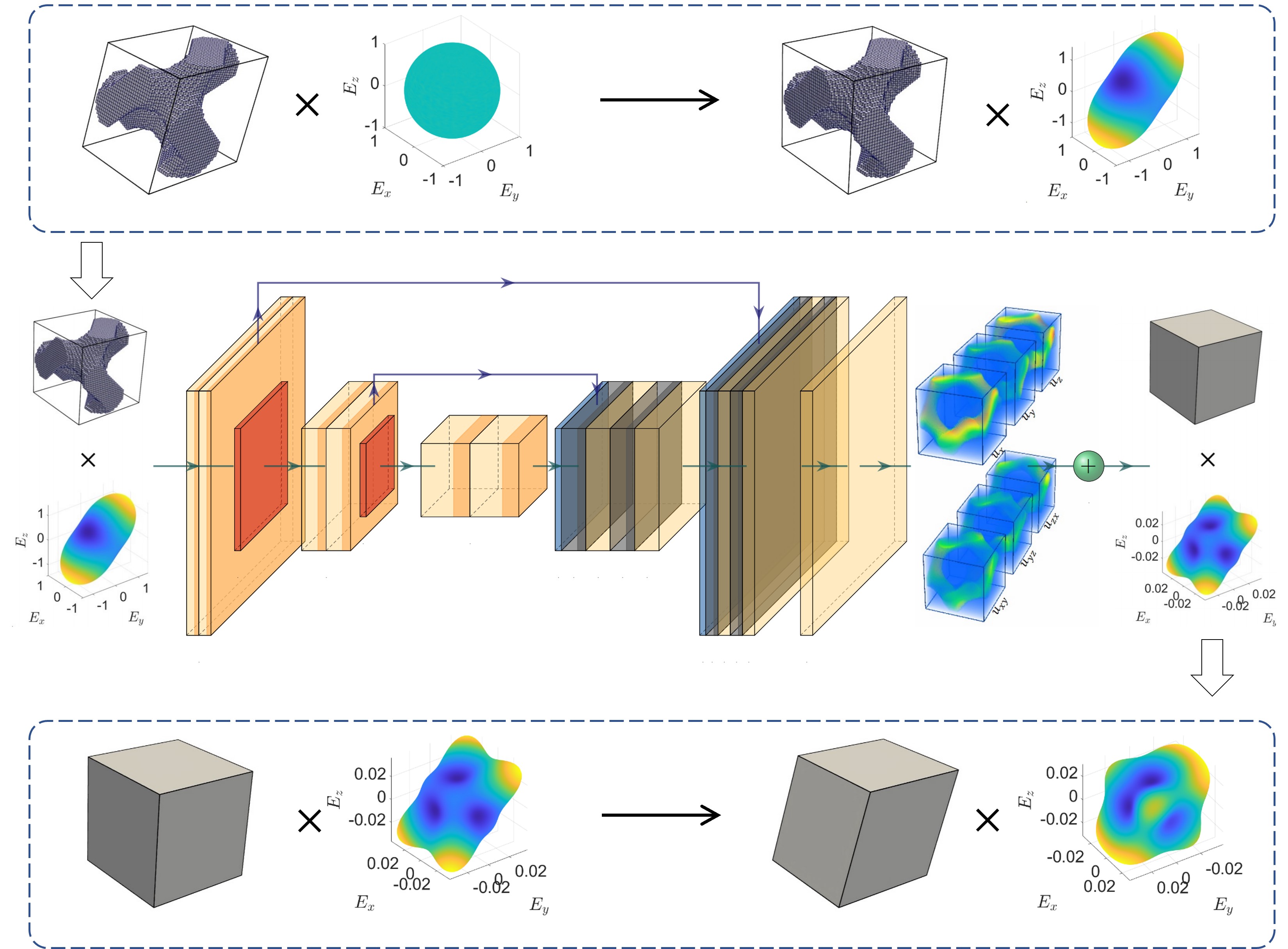}
    \put(3,57){\scriptsize (a)}
    \put(33,56.5){\scriptsize $C^b$}
    \put(87,56.5){\scriptsize $\hat{C}^b$}
    \put(3,22){\scriptsize (b)}
    \put(33,22){\scriptsize PH-Net}
    \put(3,1){\scriptsize (c)}
    \put(33,0.5){\scriptsize $\hat{C}^H$}
    \put(87,0.5){\scriptsize $C^H$}
    \end{overpic}
    \caption{The whole pipeline of PH-Net includes three main steps: (a) the pre-processing to encode microstructure, base material and boundary shape into a material-voxel tensor as input, 
    (b) predicting microscope displacements and then homogenized material properties through a convolutional neural network, (c) a post-processing step to recover the homogeneous material of parallelepiped microstructure.}
    \label{fig:overview}
\end{figure*}
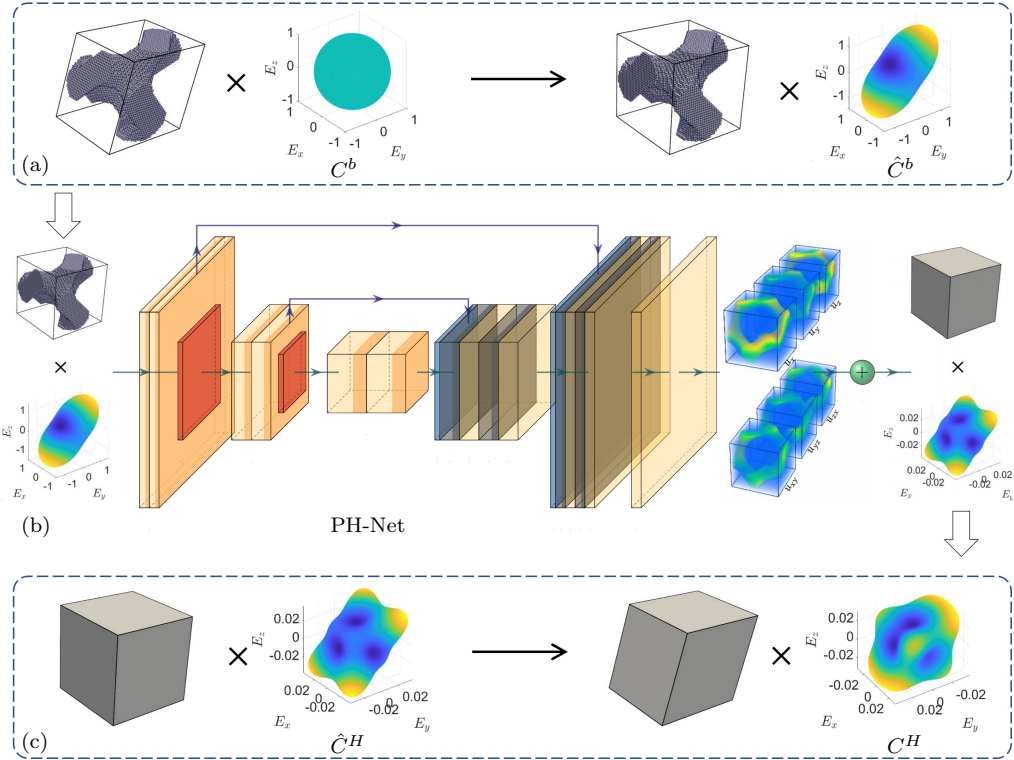

As shown in Figure~\ref{fig:overview}, the workflow of PH-Net is divided into three main stages, including (a) the pre-processing to encode microstructure, base material, and boundary shape into a material-voxel tensor as input, (b) a convolutional neural network to predict microscope displacements and then obtain homogenized material properties, (c) a post-processing step to recover the homogeneous material of parallelepiped microstructure.
In this paper, we first introduce the preliminary knowledge of the homogenization method in Section~\ref{sec:premininary_knowledge}. The input and dataset construction, as well as post-processing, are described in Section~\ref{sec:input}. Then the architecture of PH-Net and a \ph{label-free} loss function is introduced in Section~\ref{sec:architecture}.

\subsection{Preliminaries of homogenization method}
\label{sec:premininary_knowledge}
Homogenization method is used to solve general boundary value problems with periodic boundary conditions. The homogeneous material tensor $C^H_{ijkl}$ of a microstructure in a periodic unit cell $\Omega$ can be formulated as
\begin{equation}
    C^H_{ijkl}=\dfrac{1}{|\Omega|}\int_{\Omega}C^{b}_{ijkl}(\bar{\varepsilon}_{ij}-\varepsilon_{ij}(u))(\bar{\varepsilon}_{kl}-\varepsilon_{kl}(u))\mathrm{d}\Omega,
    \label{eq:continuous_integration}
\end{equation}
where $C^b$ is the base material of microstructure, $|\Omega|$ is the boundary volume of the unit cell, $\bar{\varepsilon}$ are the prescribed macroscopic strain fields and the local varying strain fields $\varepsilon(u)$ is computed by solving the following equation according to the principle of virtual work:
\begin{equation}
    \int_{\Omega}C^b_{ijkl}\varepsilon_{ij}(v)\varepsilon_{kl}(u)\mathrm{d}\Omega=\int_{\Omega}C^b_{ijkl}\varepsilon_{ij}(v)\bar{\varepsilon}_{kl}\mathrm{d}\Omega,\ \forall v \in \Omega,
\label{eq:localization}
\end{equation}
where $v$ is a virtual displacement field and $u$ is the local displacement fields under the specified macroscopic strain fields $\bar{\varepsilon}$. 
For most problems, homogenization is performed numerically by discretization, solving the homogeneous material properties with the finite element method. By discretizing the periodic boundary domain $\Omega$ into finite elements $e$, we base on a theory of numerical homogenization method called asymptotic homogenization~\cite{kalamkarov1987determination,kalamkarov1992composite}, which can be implemented into two steps.
First is the localization step according to Equation~\ref{eq:localization} with a discretization form:
\begin{equation}
    \sum_{e\in\Omega}B_e^TC^bB_eu_e\mathrm{d}\Omega=\sum_{e\in\Omega}B_e^TC^b\bar{\varepsilon}\mathrm{d}\Omega,
    \label{eq:discrete_localization}
\end{equation}
where $B_e$ is the shape matrix of element $e$ and we have $\varepsilon_e(u)=B_eu$.
With the representation of Voigt notation, the macroscopic strain fields $\bar{\varepsilon}$ can be expressed into six separated  vectors as:
\begin{equation}
\begin{aligned}
    \bar\varepsilon_1 = \{1,0,0,0,0,0\},\ \bar\varepsilon_2 = \{0,1,0,0,0,0\},\\ \bar\varepsilon_3 = \{0,0,1,0,0,0\},\  
    \bar\varepsilon_4 = \{0,0,0,1,0,0\},\\\bar\varepsilon_5 = \{0,0,0,0,1,0\},\  \bar\varepsilon_6 = \{0,0,0,0,0,1\}.
    \label{equ:macro_strain}
\end{aligned}
\end{equation}
For each element $e\in\Omega$, let $K_e=\sum\limits_{e\in\Omega}B_e^TC^bB_eu_e\mathrm{d}\Omega$ and $f_e=\sum\limits_{e\in\Omega}B_e^TC^b\bar{\varepsilon}\mathrm{d}\Omega$, then assemble them into a global stiffness matrix $K$
and force traction $f$, the corresponding microscopic local displacement fields $u$ can be computed by solving the system of linear equation $Ku=f$. 

Within the scenario of linear elasticity, the homogeneous material tensor $C^H$ is given by the integration step, in which the relevant of local strain fields $\varepsilon_e$ and macroscopic strain fields $\bar{\varepsilon}$ are in linear expression $A_e=\bar{\varepsilon}-\varepsilon_e(u_e)=\bar{\varepsilon}-B_eu_e$ for each element.
Then according to Equation~\ref{eq:continuous_integration}, the homogeneous material tensor can be integrated by
\begin{equation}
    C^H=\dfrac{1}{|\Omega|}\sum_{e\in\Omega}A^TC^b A\mathrm{d}\Omega=\dfrac{1}{|\Omega|}\sum_{e\in\Omega}(\varepsilon_0-\varepsilon_e(u_e))^T C^b(\varepsilon_0-\varepsilon_e(u_e))\mathrm{d}\Omega.
    \label{eq:discrete_integration}
\end{equation}

\subsection{Input and dataset generation}
\label{sec:input}
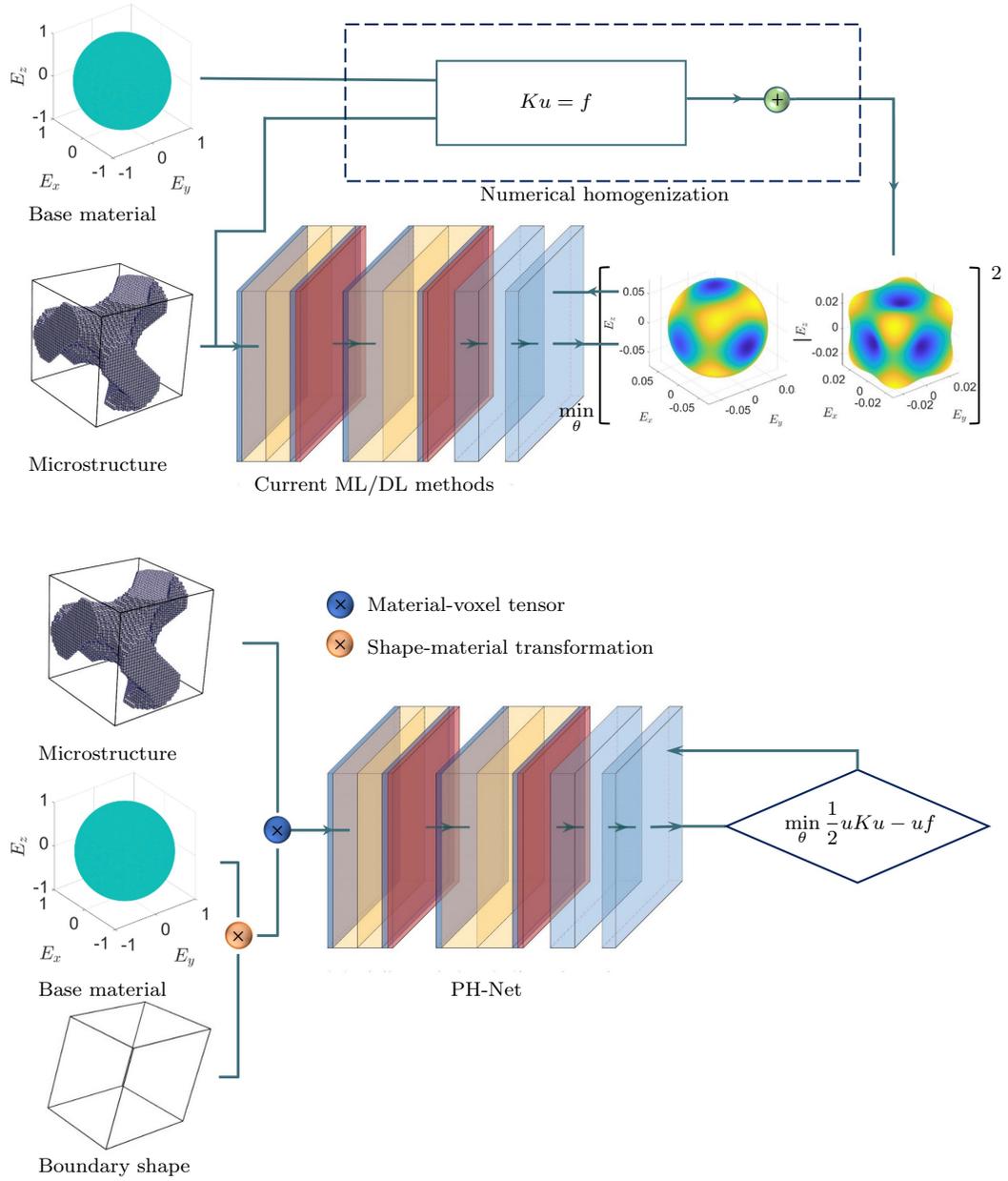
\begin{figure}
    \centering
    \begin{overpic}[width=\linewidth,tics=10]{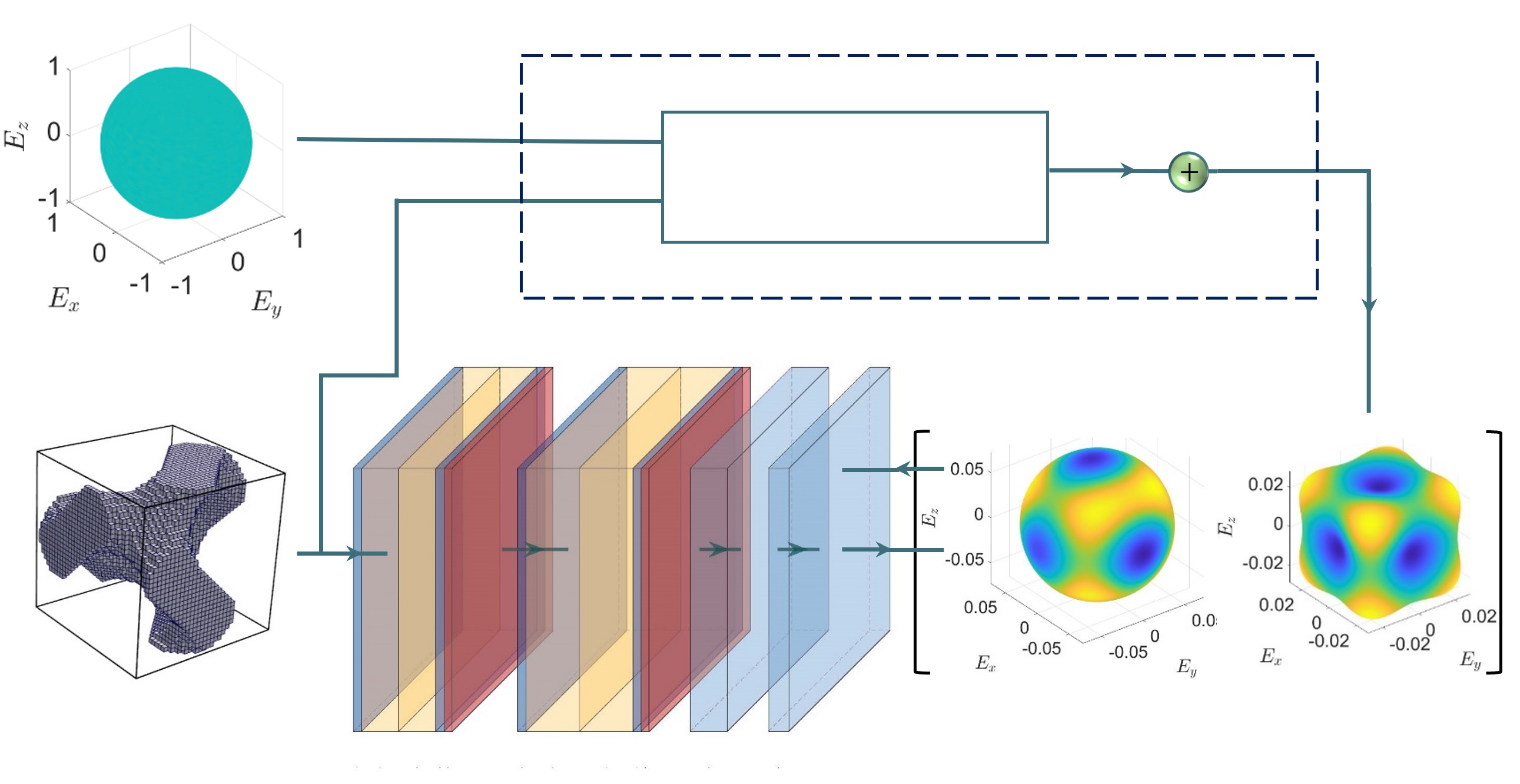}
    
    \put(2,2.5){\scriptsize {Microstructure}}
    \put(2,28){\scriptsize {Base material}}
    \put(48,30){\scriptsize {Numerical homogenization}}
    \put(52,39.5){\scriptsize{$Ku=f$}}
    \put(25,0.5){\scriptsize Current ML/DL methods}
    \put(56,8){\scriptsize $\min\limits_{\theta}$}
    \put(100,22){\scriptsize $2$}
    \put(78.6,15){ $-$}
    \end{overpic}
    \newline
    \newline
    \begin{overpic}[width=\linewidth,tics=10]{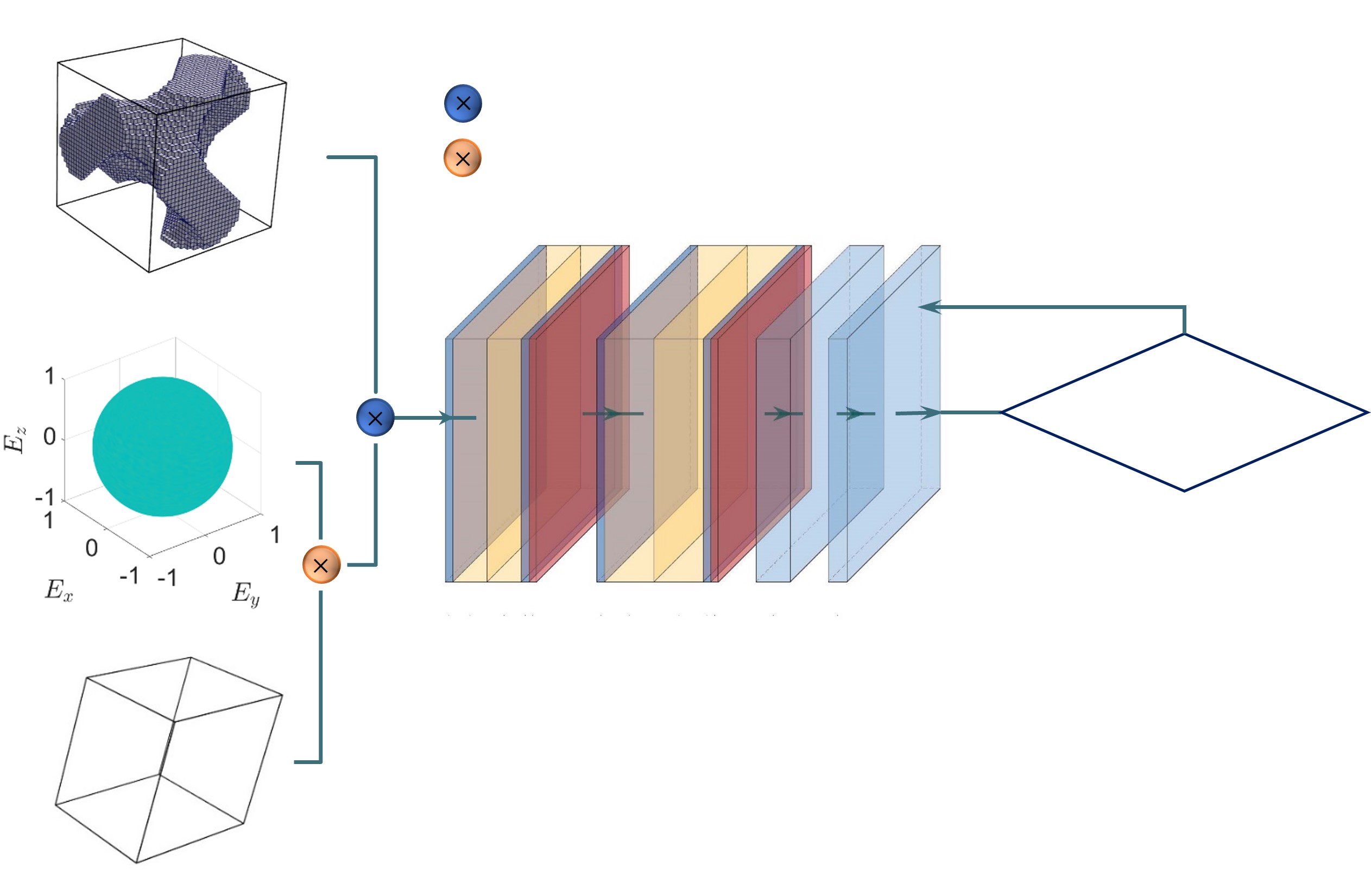}
     \put(3,42){\scriptsize {Microstructure}}
    \put(3,18){\scriptsize {Base material}}
    \put(3,0){\scriptsize {Boundary shape }}
    \put(36.5,57.2){\scriptsize {Material-voxel tensor }}
    \put(36.5,52.8){\scriptsize {Shape-material transformation }}
    \put(79,35){\scriptsize$\min\limits_{\theta}\dfrac{1}{2}uKu-uf$}
    \put(45,18){\scriptsize PH-Net}
    \end{overpic}
    \caption{The comparison of input between current ML/DL homogenization predictor (top) and PH-Net (bottom).}
    \label{fig:input_comaprison}
\end{figure}

    
    

Unlike current ML/DL methods (See Figure~\ref{fig:input_comaprison} top), which only consider a voxel-based microstructure as the input, PH-Net (Figure~\ref{fig:input_comaprison} bottom) performs the input in a representation of a material-voxel tensor, in which \textit{microstructure, its base material, and corresponding boundary shape} are taken into account. 
The main step of input generation in PH-Net is constructing the material-voxel tensor (blue operation), which benefits PH-Net training with different base materials. Then for parallelepiped microstructures, we introduce a shape-material transformation (yellow operation) to transfer the change of microstructure boundary shape into the change of its base material.

\subsubsection{Material-voxel tensor}
For a given microstructure $\Omega$ in the dataset, regardless of its geometric parameters, it is voxelized into a 0-1 3D tensor with the size $N=n^3$, where $n$ is the voxel grid resolution. It is divided into two parts, of which the solid part is 1. To be specific, the usage of our approach is limited to linear elasticity. Therefore, we implement isotropic material $C^b(E,v)$ as the base material in the dataset, which can be easily calculated by Young's modulus $E$ and Poisson ratio $v$.

However, CNN operations, i.e., \textit{conv} and \textit{pooling}, require the input of PH-Net to be a dense tensor. Thus we assume that the microstructure in the dataset is composed of a two-phase composite of hard material $C^{bh}$ and soft material $C^{bs}$, in which Young's modulus of soft material $E^s$ is much smaller than $E^h$ (i.e., $E^s=10^{-6}E^h$). 
For each input to PH-Net, the base materials $\hat C^{bh}$ and $\hat C^{bs}$ ($6\times6$ elastic tensor in Voigt notation) are reshaped into a vector, respectively. Then the material-voxel tensor is computed by $C^{bh}\times\Omega+C^{bs}\times(1-\Omega)$, such base material is embedded with voxel matrix.

\begin{figure}
    \centering
    \includegraphics[width=.7\linewidth]{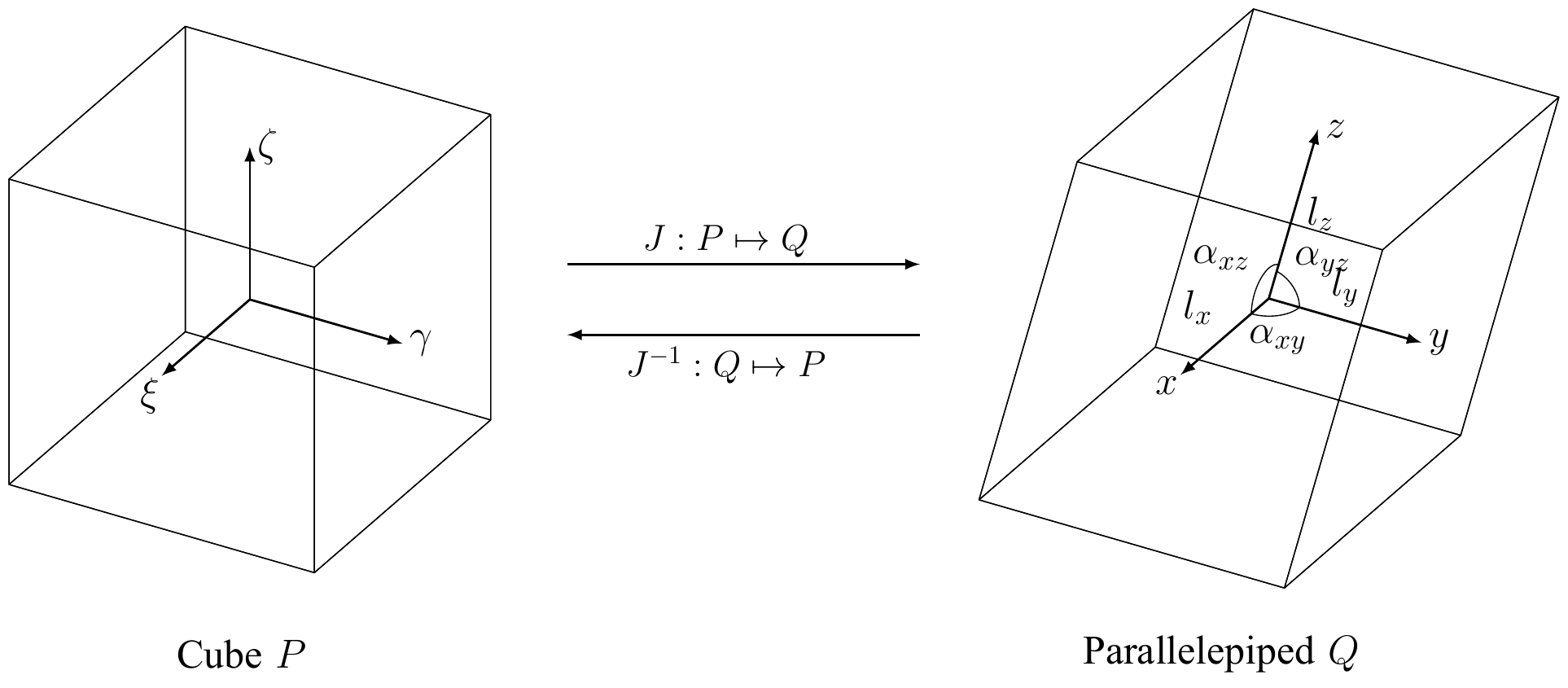}
    \caption{The shape transformation $J$ and its inverse between cube $P$ and parallelepiped $Q$ are considered as the transformation of coordinate basis with three scaling variables $l_x,l_y,l_z$ and three angle variables $\alpha_{xy},\alpha_{yz},\alpha_{zx}$, respectively.}
    \label{fig:parameters}
\end{figure}

\subsubsection{Shape-material transformation}

It is known that the shape transformation from a cube $P$ to a parallelepipe $Q$ can be considered as an axes transformation changes from an orthogonal coordinate system $(\xi,\eta,\zeta)$ to a nonorthogonal basis $(x,y,z)$.
Suppose an invertible matrix $J:P\mapsto Q$ denotes a transformation of axes from $P$ to $Q$.
As shown in Figure~\ref{fig:parameters}, the affine transformation $J:P\mapsto Q$ can be parameterized by 6 variables, which are the magnitude scales of coordinate axes and the angles between them.
We use $l_x, l_y, l_z$ and $\alpha_{xy}, \alpha_{yz}, \alpha_{xz}$ to denote the scale variables and angle variables, respectively. 
To eliminate rotational errors, we specify that the X-axis of the deformed hexahedron is $[1,0,0]^T$, and the $X-Y$ plane is perpendicular to $[0,0,1]^T$.
Therefore, we can express the mapping relationship as the product of the scaling matrix $T(l_x,l_y,l_z)$ and the shear matrix $S(\alpha_{xy}, \alpha_{yz}, \alpha_{xz})$, that is $J = ST$. 
Similarly, we can calculate linear mapping $J^{-1}:Q\to P$ by inversing $J$ in turn.

As we know, there has not been a DL homogenization predictor implemented to predict homogeneous material of parallelepiped microstructures at present. Most DL methods adopt voxel-based binary inputs for their innately fit for DL operations, like \textit{conv} and \textit{pooling}.  However, the position information for voxel matrix is intrinsically embedded and not intuitive to couple boundary shape and binary voxel matrix.
To address this problem, we adopt a shape-material transformation introduced by \cite{low-parametric}. 
For a microstructure inside the boundary volume of $P$ and $Q$, if we have
\begin{equation}
    \varepsilon_{ij}=J_{ik}J_{jl}\hat\varepsilon_{kl},\ \sigma_{ij}=J_{ik}J_{jl}\hat\sigma_{kl},
    \label{eq: shape_material_transformation_1}
\end{equation}
where the second-order tensors $\hat{\varepsilon}, \hat{\sigma}$  and $\varepsilon, \sigma$  are defined as the microscope strains and stresses of the microstructure inside a cube $P$ and a parallelepiped $Q$, respectively. Then the shape-material transformation can be described as:~\\~\\
\textbf{Proposition:}
\textit{Let $\hat{C}^H=\hat{\mathcal{H}}(\hat{C}^b,P)$ and $C^H=\mathcal{H}(C^b,Q)$  denote homogenization computation in the boundary volume of cube $P$ and parallelepipe $Q$, respectively. If we perform a shape-material transformation
\begin{equation}
    \hat{C}^b_{ijkl}=J^{-1}_{pi}J^{-1}_{qj}J^{-1}_{rk}J^{-1}_{sl}C^b_{pqrs}, 
\end{equation}
then homogeneous material properties in the parallelepiped domain $Q$ can also be computed by shape-material transformation
\begin{equation}
     C^H_{ijkl}=J_{pi}J_{qj}J_{rk}J_{sl}\hat{C}^H_{pqrs}.
    \label{eq: shape_material_transformation_2}
\end{equation}
}
Note the shape-material transformation holds only if $Q$ is a parallelepiped, more details of the proof of shape-material transformation can be found in the appendix.
Therefore, the shape-material transformation operation of PH-Net works as follows:
\begin{enumerate}
    \item Compute $\hat{C}^{bh}$ and $\hat{C}^{bs}$ by Equation~\ref{eq: shape_material_transformation_1}, then construct $\hat C^{bh}\times\Omega+\hat C^{bs}\times(1-\Omega)$ as input (Figure~\ref{fig:overview}a);
    \item Predict microscope displacements field $\hat u$ in $\hat{\mathcal{H}}(\hat{C}^b,P)$ using PH-Net, and integrate homogeneous material $\hat{C}^H$ (Figure~\ref{fig:overview}b);
    \item Recover homogenized tensor $C^H$ on $Q$ by Equation~\ref{eq: shape_material_transformation_2} (Figure~\ref{fig:overview}c).
\end{enumerate}

\subsection{Architecture of PH-Net}
\label{sec:architecture}

Unlike other ML/DL methods that predict homogeneous properties directly, PH-Net predicts the microscopic displacement fields $u$ from given macroscope strains $\bar\varepsilon$ in the localization step through a 3D convolutional neural network $\mathbb{N}(\theta)$. 
The main difference is that we construct a novel loss function according to the minimum potential energy (MPE) theory instead of the mean squared (MSE) loss function.
Our loss function is formulated as
\begin{equation}
    L(u) = \sum_{i=1}^6\frac{1}{2} u^T_i K u_i- u^Tf_i,
    \label{loss}
\end{equation}
whose gradients $\dfrac{\partial L}{\partial u}$ are expressed as $\sum\limits_{i=1}^6 Ku_i-f_i$. 
This loss function is reasonable and has been commonly used in many works~\cite[e.g.][]{Samaniego_2020}  that solve finite element (FE) problems by ML and DL methods.
The macroscope strains $\hat\varepsilon$ can be regarded as universal and implicated ground-truth in the loss function of PH-Net.
That means our energy-based loss function as a criterion is enough to minimize PH-Net to convergence.
The MPE-based loss function benefits PH-Net to be label-free, avoiding the most time-consuming part- building ground-truth via numerical homogenization, compared with previous data-driven methods. This part cannot be neglected, especially for parallelepiped microstructures. 
Besides, PH-Net is formulated to predict the solution of a PDE problem instead of homogeneous properties directly, which gains more potential with the generalization of different microstructure types, base materials, and boundary shapes.

We construct PH-Net with a U-Net style structure, as shown in Figure~\ref{fig:overview}(b). 
The input of PH-Net is a fourth-order tensor with a size $36\times n^3$, in which the boundary shape, microstructure morphology, and its base materials are in consideration. 
The output of PH-Net is $18\times n^3$ in size, which is spitted into six local displacement fields under the load of six macroscopic strain fields $\bar\varepsilon$.
%
Compared with the fully connected network (FCN), U-Net~\cite{ronneberger2015u} implements an up-sampling operation to replace the pooling operation and adds a jump connection module with a more elegant network architecture. 
In practice, we observe that using up-sampling and down-sampling operations on microstructure space benefits the connection and the exchanges of weight parameters, so U-Net can be localized and achieve better results with only a small number of training data. 

As illustrated in Figure~\ref{fig:overview}(b), PH-Net contains a contracting path (left side) and an expansive path (right side) that is symmetrically distributed at the beginning and end of the network. 
The contracting path contains two repeated convolution blocks, each consisting of a 3D convolutional layer and a rectified linear unit (ReLU). A batch-norm operation is performed at the end of each convolutional layer to reduce the impact of data distribution. The convolution kernel is $3\times 3\times 3$, and padding is set to 1. The convolution block is followed by a $2\times 2\times 2$ maximum pooling layer with a stride 2 for down-sampling. At each down-sampling step, we double the number of feature channels. 
On the expansive path, the feature map clipped from the contracting path is connected with the up-sampled results, and then the convolution operation is carried out. Different from the compression path, the number of channels is halved after each convolution. Finally, we use a $1\times 1\times 1$ convolution layer to map the result to an $18\times n^3$ tensor.

\section{Results and Discussions}
\label{sec:results}

\subsection{Setting of PH-Net}
In this paper, we construct two main datasets to test the generalization performance of PH-Net in terms of different microstructure types, base materials, and boundary shapes.
All datasets are shuffled and partitioned into two parts: $80\%$ of the samples are selected as the training set, and the remaining $20\%$ as the testing set to evaluate the prediction error of PH-Net.

\textbf{Dataset 1} is designed to test the generalization of PH-Net in terms of different base materials and boundary shapes. In this dataset, we choose a triply periodic minimal surface (TPMS) called \textit{Tubular Gyroid} (TG) as the microstructure with 40 uniform samples in volume fraction $[2\%, 33\%]$.
We randomly select $1500$ distinct boundary shapes for each volume fraction sample in range of shape parameters $\alpha_{xy},\alpha_{yz},\alpha_{xz}\in[75^\circ,90^\circ]$ and $l_x,l_y,l_z\in[1,2]$ and hence we have 60K samples in total. To ensure consistency of input, we normalize all samples to the same boundary volume. 
The hard base material $C^{bh}$ and soft base material $C^{bs}$ are set to $E^{h}=1,\ v=0.3$ and $E^s=1\times 10^{-6},v=0.3$, respectively. Due to shape-material transformation, the transformed base material of each sample is different with others.

\textbf{Dataset 2} is designed to test the generalization of PH-Net w.r.t different microstructure types. Considering huge topological differences, we divide this dataset into two sub-datasets; one consists of 100 types of truss microstructures, and the other includes 16 types of shell microstructures, in which each microstructure has 40 uniform volume fraction samples. All microstructures are in cubic boundary shape, and the hard and soft base materials are the same as Dataset 1.

\begin{table}
\centering
\scriptsize
\caption{The main components of PH-Net}
\begin{tabular}{cl} \toprule
Number & Component \\ \hline
1 & Conv Block (36, 64) \\
2 & Conv Block (64, 128) \\
3 & Conv Block (128, 256) \\
4 & Deconv Block (256, 128) \\
5 & Deconv Block (128, 64) \\
6 & Deconv Block (64, 32) \\
7 & Conv (32, 18) \\ \bottomrule
\end{tabular}
\label{tab:architecture}
\end{table}
We conduct a set of hyper-parameters (e.g., batch-norm, learning rate, architecture) to modulate optimal networks. 
We design a U-Net CNN architecture for PH-Net as demonstrated in Figure~\ref{fig:overview}. The main components of PH-Net are listed in Table~\ref{tab:architecture}, in which the convolutional block is a sequence of a max pooling layer, two convolutional layers, a ReLU activation layer, and a batch normalization layer. While the de-convolutional block is similar to the convolutional block, except the max pooling layer is replaced with an upsampling layer. 
To balance the GPU memory cost and performance, the resolution of PH-Net is set to $n=48$ in our devices.
PH-Net is built on Pytorch 1.8.1 using Python 3.9, trained on a platform equipped with NVIDIA GeForce GTX 1080Ti GPU with Intel Core i3-7980CPU@2.6GHz.
For the training procedure, we implement Adam~\citep{kingma2017adam} as the optimizer of PH-Net, whose learning rate is set to $5\times10^{-4}$ and the batch size is set to 8.
The performance of PH-Net is evaluated by the error between predicted results and their corresponding ground-truth $C^H_g$, given by
\begin{equation}
    \delta={|C^H-\bar C^H|}/{|\bar C^H|},
    \label{equ:error}
\end{equation}
where $\bar C^H$ is computed by a MATLAB version homogenization method~\citep{dong2019149}.

\subsection{The performance of PH-Net}

\begin{figure}
    \centering
    \includegraphics[width=.5\linewidth]{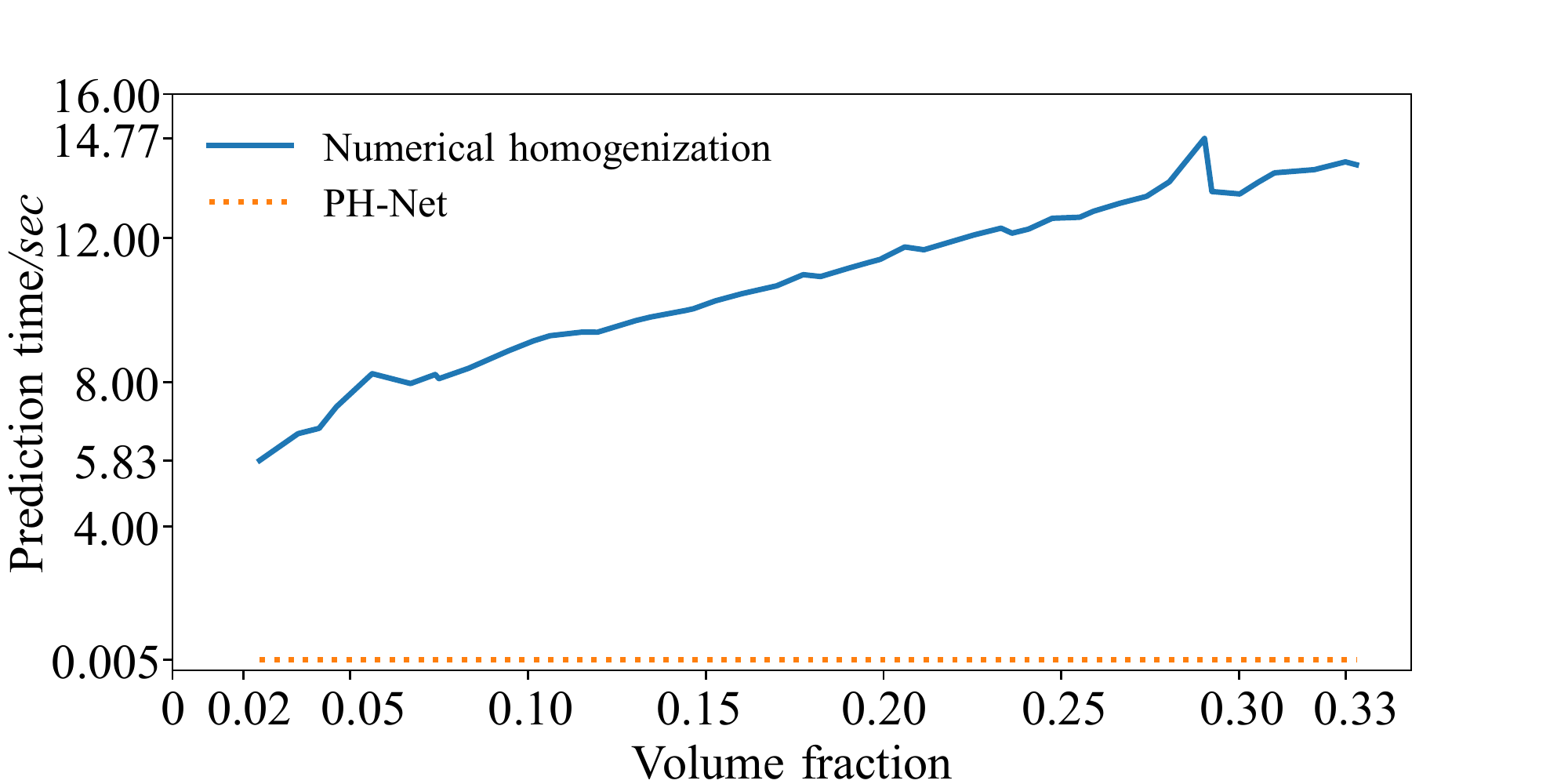}
    \caption{The prediction time of numerical homogenization and PH-Net for microstructures with different volume fractions.}
    \label{fig:volume_time}
\end{figure}

\subsubsection{Time-consuming analysis}
The training process takes 20 hours with 66 epochs on Dataset 1, of which 30\% of the time is spent evaluating prediction errors.
However, current DL prediction methods with the MSE loss function need at least 180 hours of extra time to compute the ground-truth for Dataset 1. 
We compare the prediction time of PH-Net with the numerical homogenization method, as shown in Figure~\ref{fig:volume_time}.
The time cost for numerical homogenization is positively correlated with the volume fraction of the microstructure.
Here, for the model with a volume fraction from 2\% to 33\%, the solving time increases from $5.83\mathrm s$ to $14.77\mathrm s$.
On the contrary, PH-Net gains hundreds of speed-up independent of the volume fraction, which takes $5.3\mathrm{ms}$ for all volume fractions.

\begin{figure*}
     \centering
     \begin{subfigure}[b]{0.48\textwidth}
         \centering
         \includegraphics[width=\textwidth]{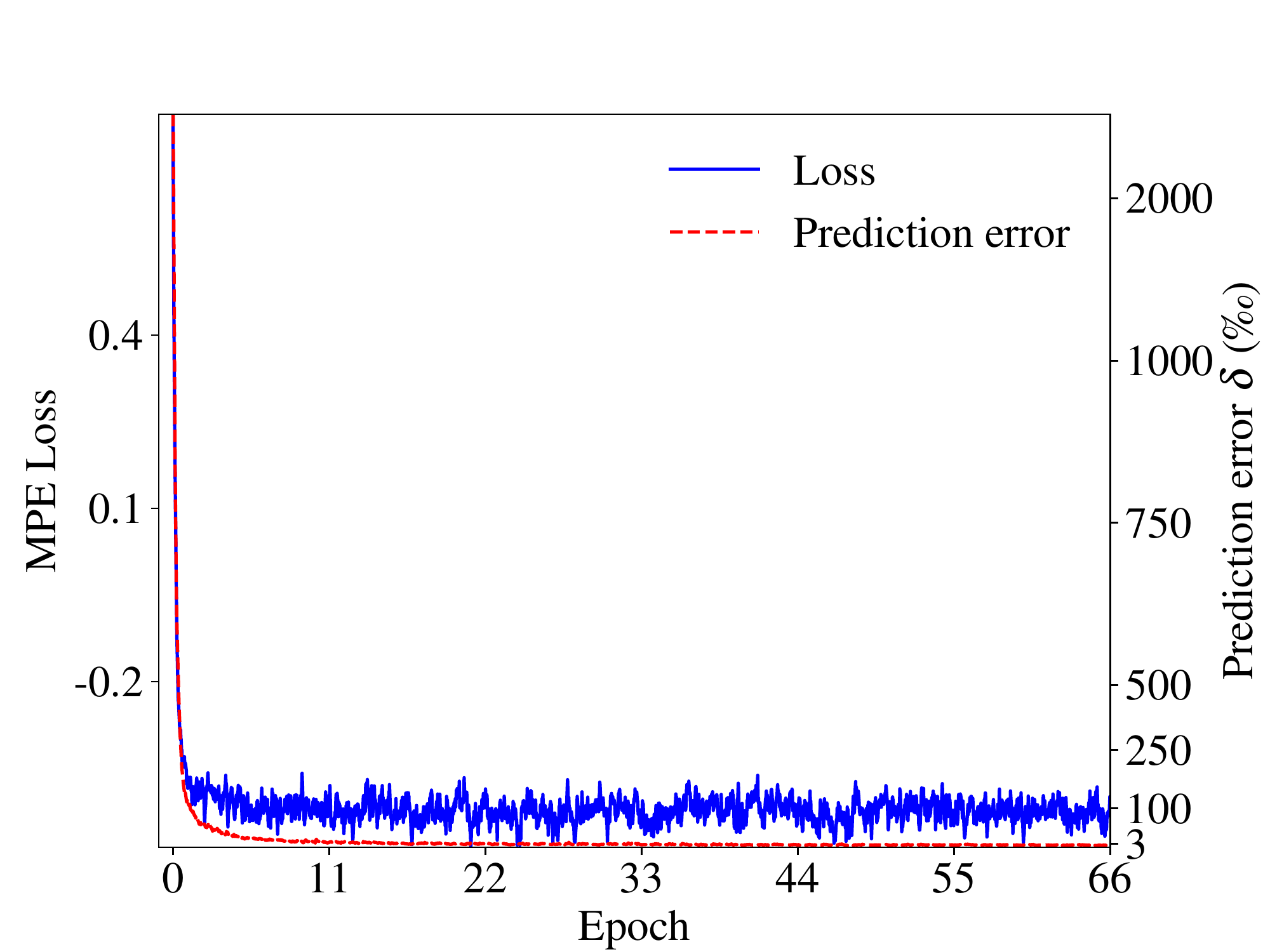}
         \label{fig:loss_error}
     \end{subfigure}
     \hspace{.02\textwidth}
     \begin{subfigure}[b]{0.48\textwidth}
         \centering
         \includegraphics[width=\textwidth]{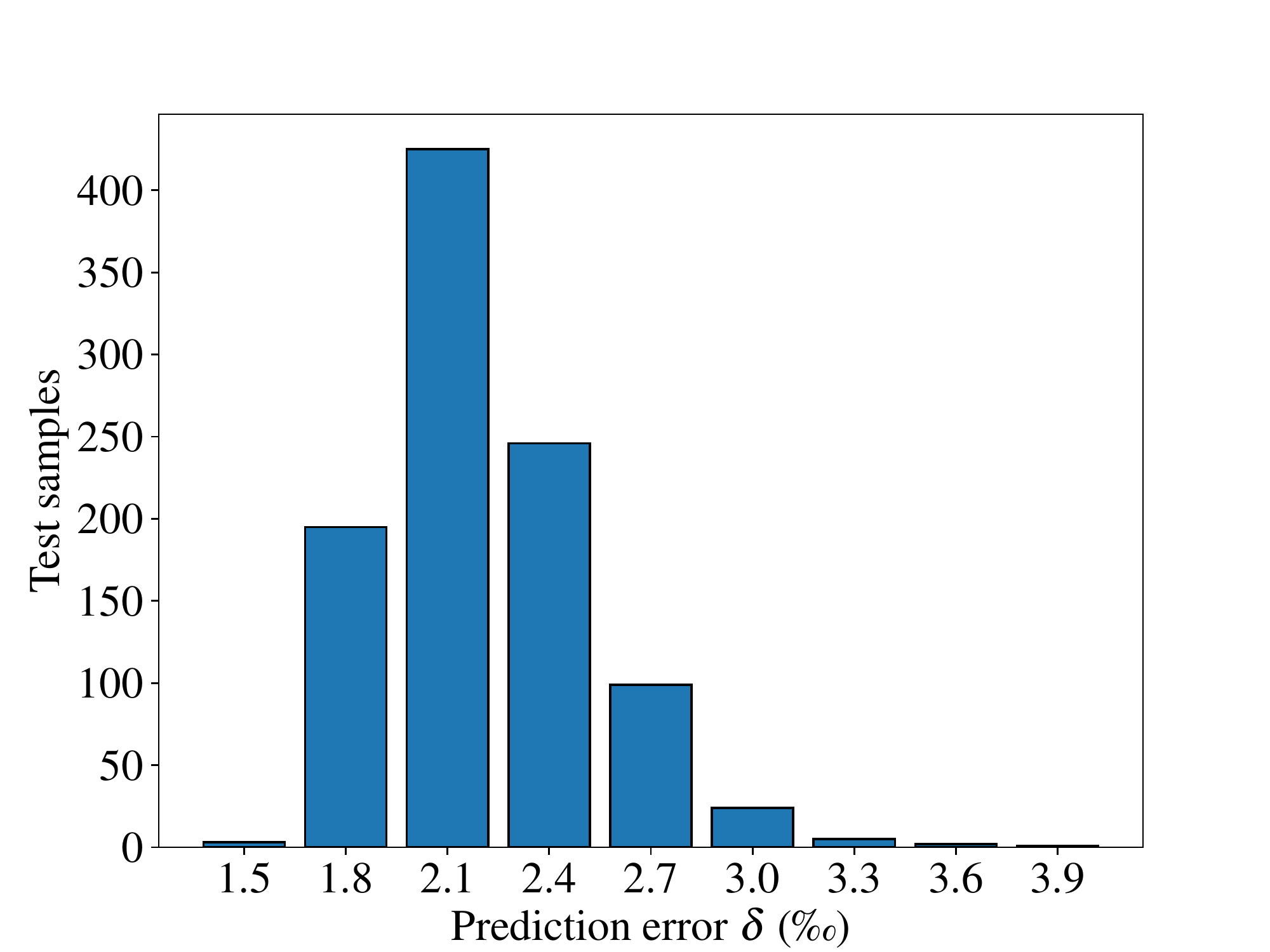}
         \label{fig:error_distribution}
     \end{subfigure}
\caption{Left: trends of loss and prediction error in the training process of PH-Net on Dataset 1; right: prediction error distribution for the testing set of Dataset 1.}
\label{fig:three graphs}
\end{figure*}

\begin{figure*}
     \centering
     \begin{subfigure}[b]{0.48\textwidth}
         \centering
         \includegraphics[width=\textwidth]{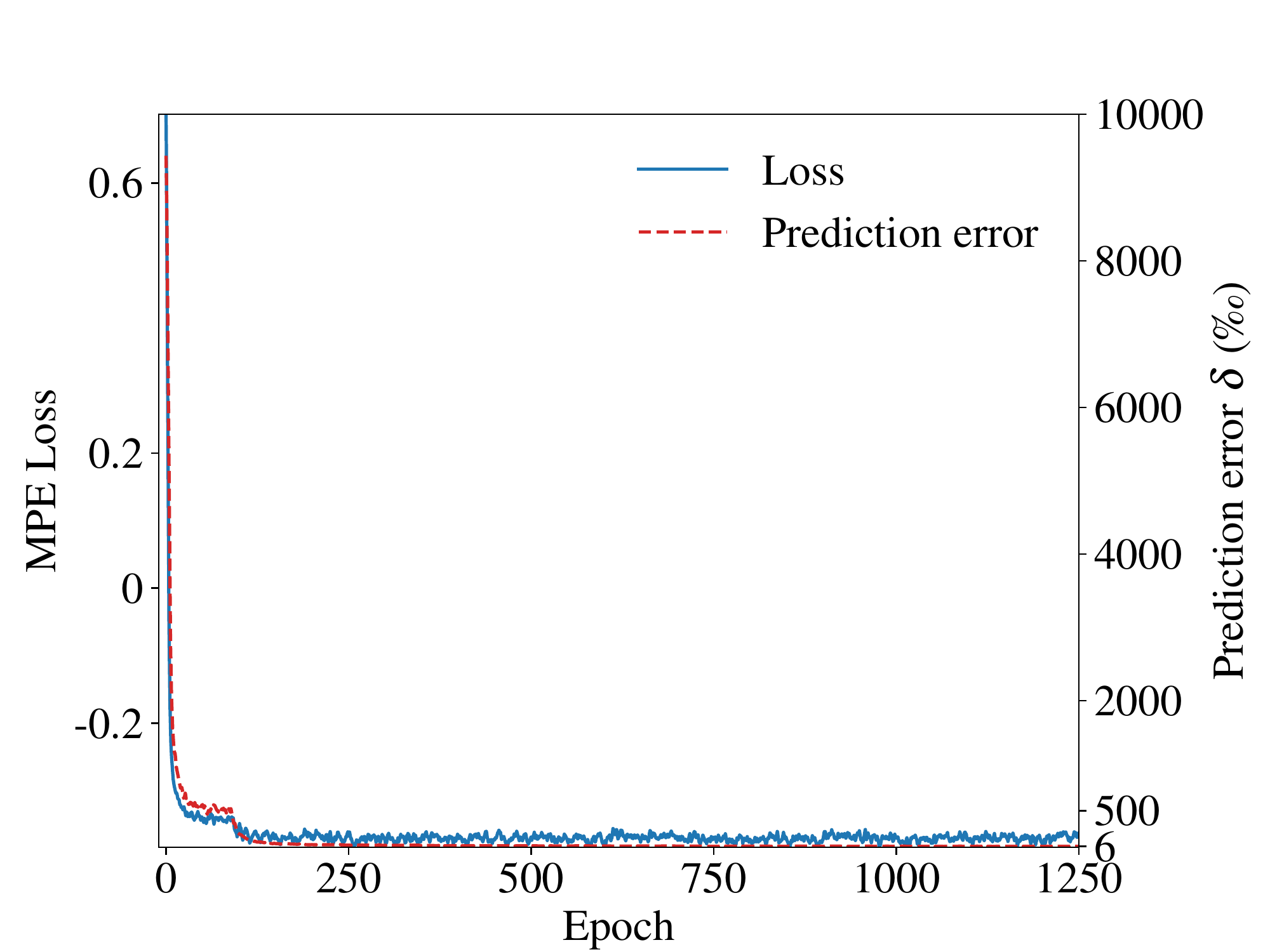}
         \label{fig:loss_error}
     \end{subfigure}
     \hspace{.02\textwidth}
     \begin{subfigure}[b]{0.48\textwidth}
         \centering
         \includegraphics[width=\textwidth]{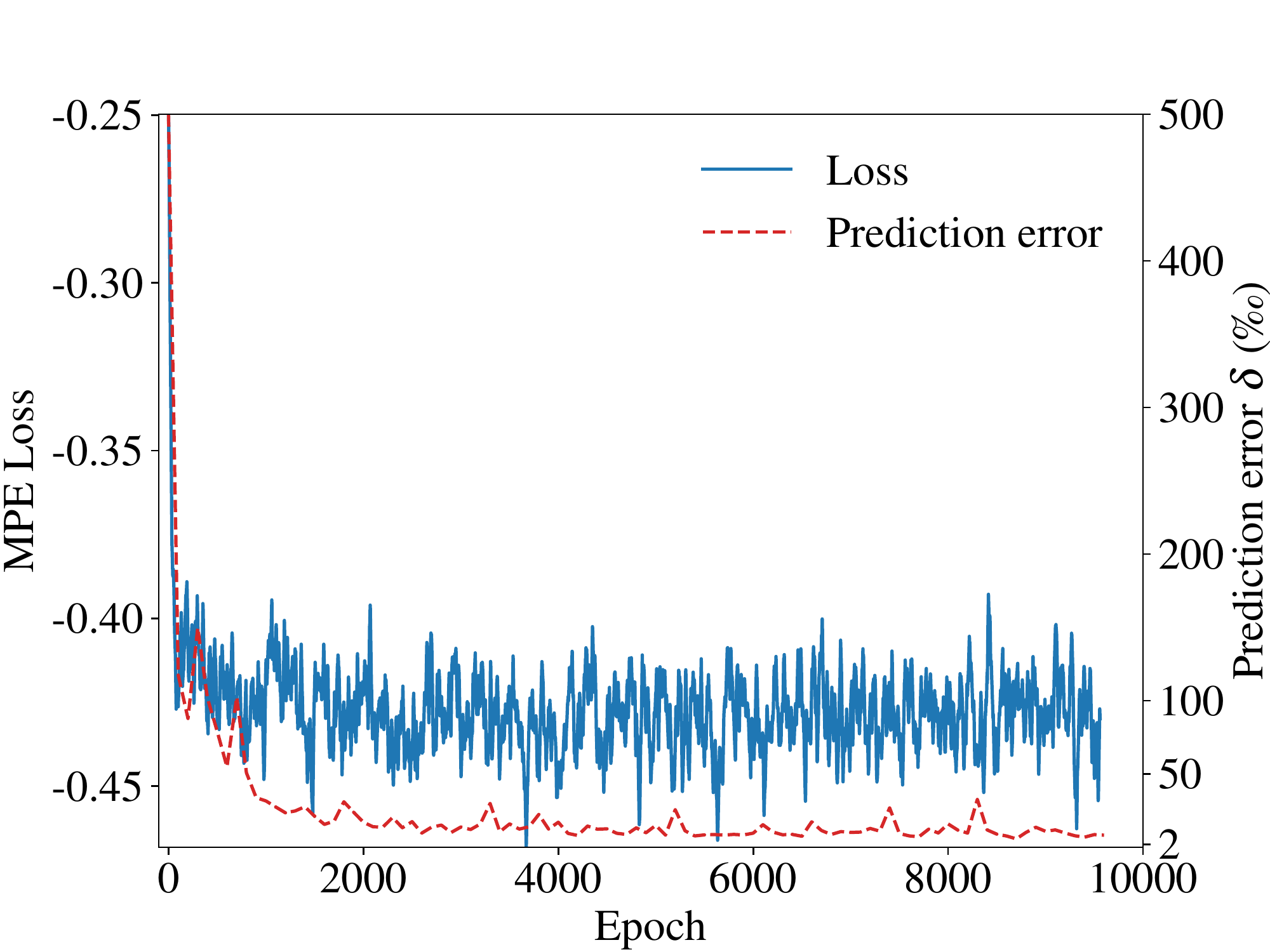}
         \label{fig:error_distribution}
     \end{subfigure}
\caption{Left: trends of loss and prediction error in the training process of truss microstructures in Dataset 2; right: trends of loss and prediction error in the training process of shell microstructures in Dataset 2.}
\label{fig:rod_shell}
\end{figure*}

\subsubsection{Convergence and error distribution}
As shown in Figure~\ref{fig:three graphs} (left), the loss of PH-Net on \textbf{Dataset 1} tends to converge after 66 epochs. 
We take one batch out of 100 batches to evaluate the prediction error during the training process, and the plot of prediction error shows a similar downward trend with loss. 
Both of them witness a rapid decline at first, then decrease slowly and fall to their minimal point after 66 epochs, where the average prediction error remains stable at 3\textperthousand. 

After the training process, the testing set is used to validate the performance of trained PH-Net, and the distribution of prediction errors is plotted in Figure~\ref{fig:three graphs} (right).
 The test results are similar to the prediction error of the last epoch of the training process, where the average prediction error for the whole test set is $2.4$\textperthousand.
The prediction error of PH-Net has achieved the same level of SOTA results after 66 epochs and can support most applications, e.g., two-scale microstructure modeling.
Increasing epochs can further reduce the prediction error, but it may be less significant than former epochs, limited by its built-in drawbacks like floating point accuracy and interpolation error.

Figure~\ref{fig:rod_shell} demonstrates the training results of Dataset 2. It shows that PH-Net has a well convergence w.r.t. both loss and prediction error on truss microstructures (left) and shell microstructures (right) after training $31.3$ hours and $20.7$ hours, respectively. The prediction error in training processing of truss and shell microstructures converges to 6.8\textperthousand\ and 2.1\textperthousand, and their average prediction errors on the testing set are $6.7$\textperthousand\ and 2.6\textperthousand, respectively. It makes sense that the prediction error for truss microstructure is larger than shell since truss microstructures include more microstructure types than shell microstructure.
Moreover, considering that most types of microstructures with the cubic boundary shape are orthotropic, we implement an improved approach of \textit{CLM Theorem} introduced by~\cite{hu2001new} to calculate the change of homogeneous material of orthotropic microstructures when the base material is changed. 
Supposing that the base material $C^b(E^b,v^b)$ is replaced with another base material $C^b(\hat E^b,\hat v^b)$, then the orthotropic homogenized material is changed from $C^H(E_i, v_{i,j}, G_{i,j})$ to $C^H(\bar E_i, \bar v_{i,j},\bar G_{i,j}),\  i,j\in\{x,y,z\}$ by
\begin{equation}
    \begin{split}
        \bar{E}_i&=\frac{\bar{E}^b}{E^b}E_i,\\
        \bar{v}_{i,j}&=v_{i,j}-(v^b-\bar{v}^b)\frac{E_i}{E^b},\\
        \frac{1}{2 \bar G_{i,j}}&=\frac{ E^b}{\bar E^b}\frac{1}{2 G_{i,j}}-(v^b-\bar v^b).
    \end{split}
    \label{equation:replacement}
\end{equation}

\begin{figure}
\centering
\includegraphics[width=.8\linewidth]{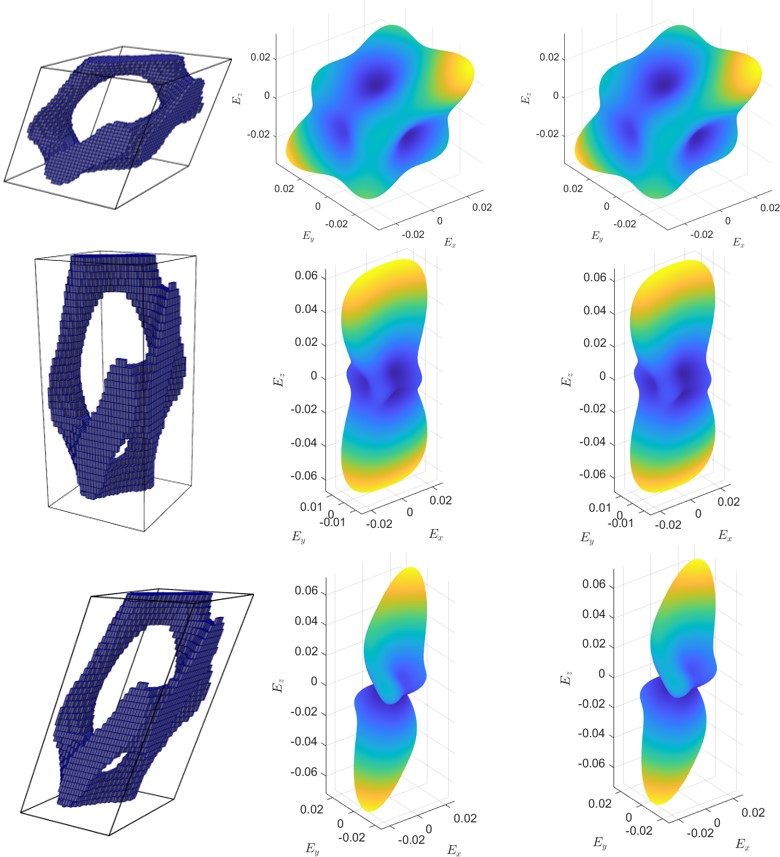}
\caption{Three parallelepiped TPMS-TG microstructures with the same volume fraction (10\%) but different shape parameters (left column). From top to bottom, their shape parameters are $(75^{\circ},75^{\circ},75^{\circ},1,1,1)$, $(90^{\circ},90^{\circ},90^{\circ},1,1,2)$, $( 75^{\circ},75^{\circ},75^{\circ},1,1,2)$, respectively.
The second and third column are the plots of predicted Young's modulus given by numerical homogenization (middle) and PH-Net (right).}
\label{fig:examples}
\end{figure}

\begin{figure}
    \centering
    \includegraphics[width=.7\linewidth]{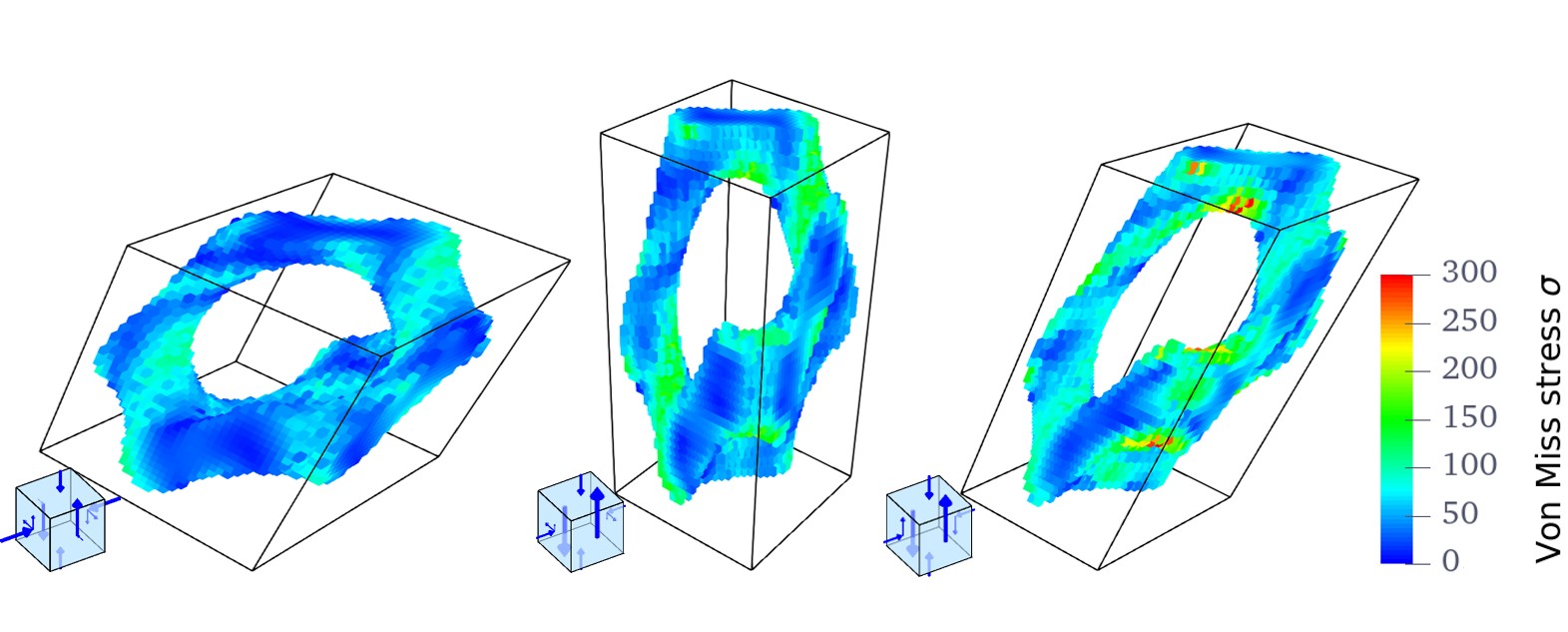}
    \centerline{\scriptsize $\sigma_{max}=146.5\mathrm{Pa}$\hspace{.08\linewidth}$\sigma_{max}=276.4\mathrm{Pa}$\hspace{.08\linewidth}$\sigma_{max}=412.2\mathrm{Pa}$\hspace{.07\linewidth}}
\caption{The worst-case stress distribution of the three examples in Figure~\ref{fig:examples}.}
\label{fig:worst_case_stress}
\end{figure}

\subsubsection{Demonstration of prediction results} 
To further evaluate the performance of PH-Net, we demonstrate examples in our dataset with more visualization results.
As shown in Figure~\ref{fig:examples}, we provide three parallelepiped TPMS-TG microstructures with the same volume fraction (10\%) but different shape parameters. 
Since Young's modulus is a crucial variable to depict the material properties, we compared their plots of Young's modulus surface $E$ with the prediction results of the numerical homogenization method and PH-Net. 
We observe that the predicted results of PH-Net (right) are incredibly close to those of the numerical homogenization method (middle column). The prediction error for the three examples is only 2.2\textperthousand, 2.8\textperthousand, 2.6\textperthousand, respectively.

Apart from predicting homogenized material properties as current DL methods, PH-Net can benefit the prediction of microscope mechanical properties, i.e., local stresses, attributing to the choice of displacement as output.
As shown in Figure~\ref{fig:worst_case_stress}, we input the predicted displacements of three parallelepiped microstructures, then obtain their worst-case stress distribution. 
Theoretically, we can calculate and visualize mechanical properties of all microstructures introduced by \cite{liu2021mechanical}, e.g., local strains and stresses, yield strength, and shear strength. 
These properties heavily relying on the prediction of displacements are much helpful in microstructure design but are not available with existing DL methods.

\subsection{Physical experiments}
\begin{figure*}
    \centering
    \includegraphics[width=\linewidth]{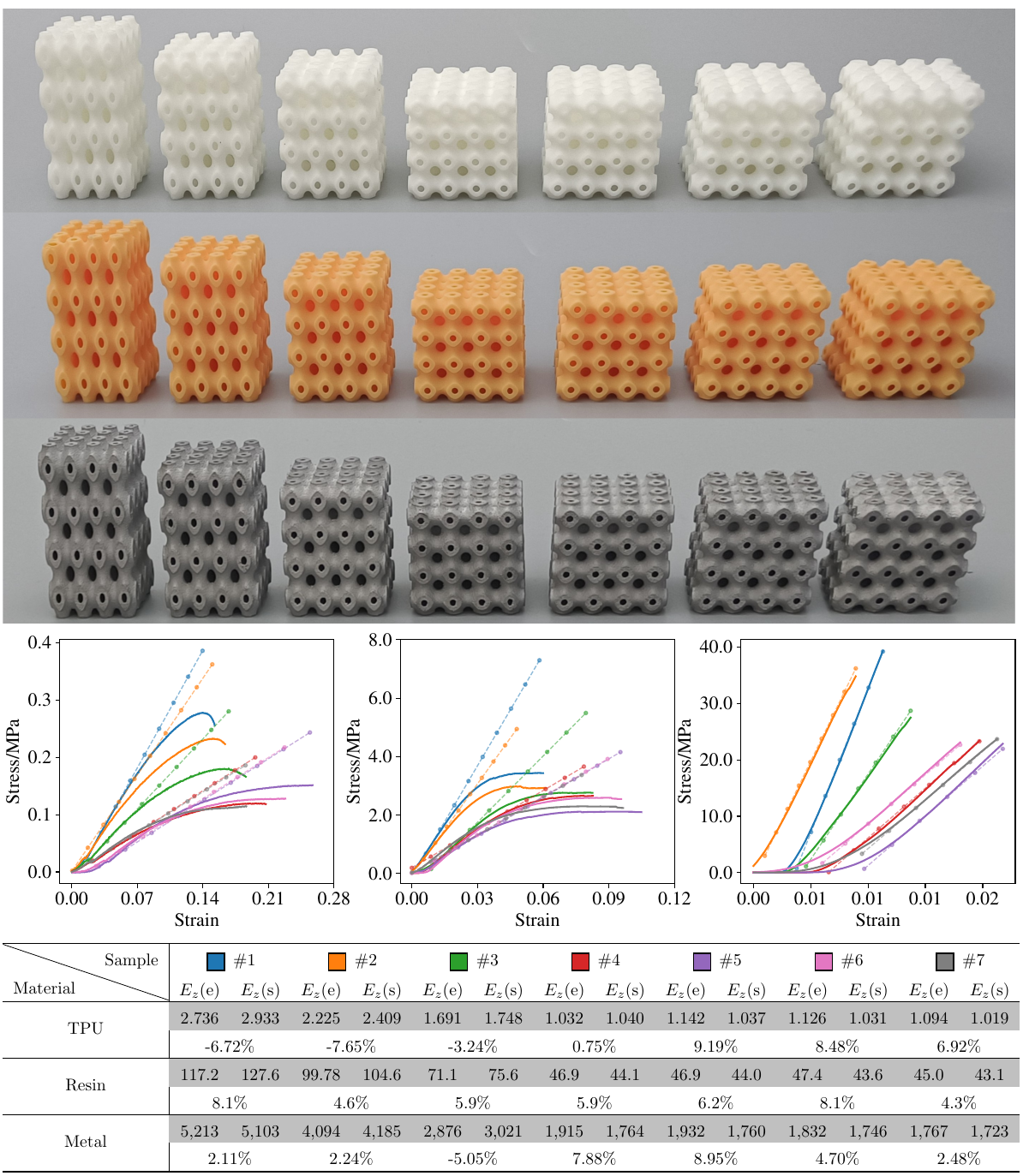}
    \caption{Physical experiments for different base materials. Top: test samples printed by TPU ($E=24\mathrm{MPa},v=0.45$), resin ($E=750\mathrm{MPa}, v=0.25$) and metal ($E=43\mathrm{GPa}, v=0.25$); middle: strain-stress curves of z-direction (solid line) and their homogeneous prediction of linear part (dashed line) for different base materials; bottom: comparison of Young's modulus of z-directions w.r.t physical experiments $E_z(e)$ and PH-Net simulation $E_z(s)$ and their relative errors for different base materials.}
    \label{fig:diff_material}
\end{figure*}

\begin{figure*}
    \centering
    \includegraphics[width=\linewidth]{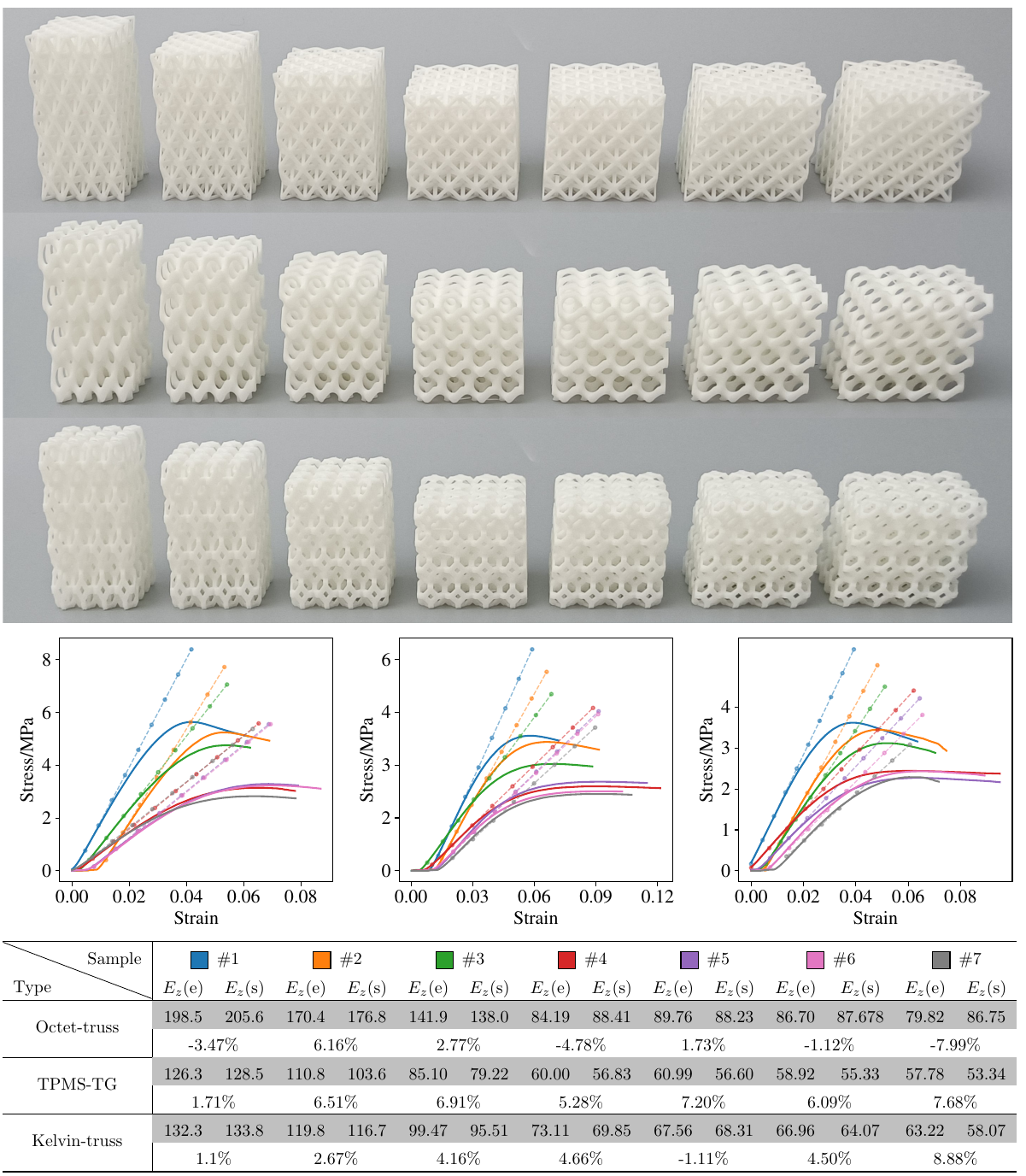}
    \caption{Physical experiments for different microstructure types. Top: test samples with type of Octet-truss, TPMS-TG and Kelvin-truss; middle: strain-stress curves of z-direction (solid line) and their homogeneous prediction of linear part (dashed line) for different microstructure types; bottom: comparison of Young's modulus of z-directions w.r.t physical experiments $E_z(e)$ and PH-Net simulation $E_z(s)$ and their relative errors for different microstructure types.}
    \label{fig:diff_type}
\end{figure*}

\begin{figure*}
    \centering
    \includegraphics[width=\linewidth]{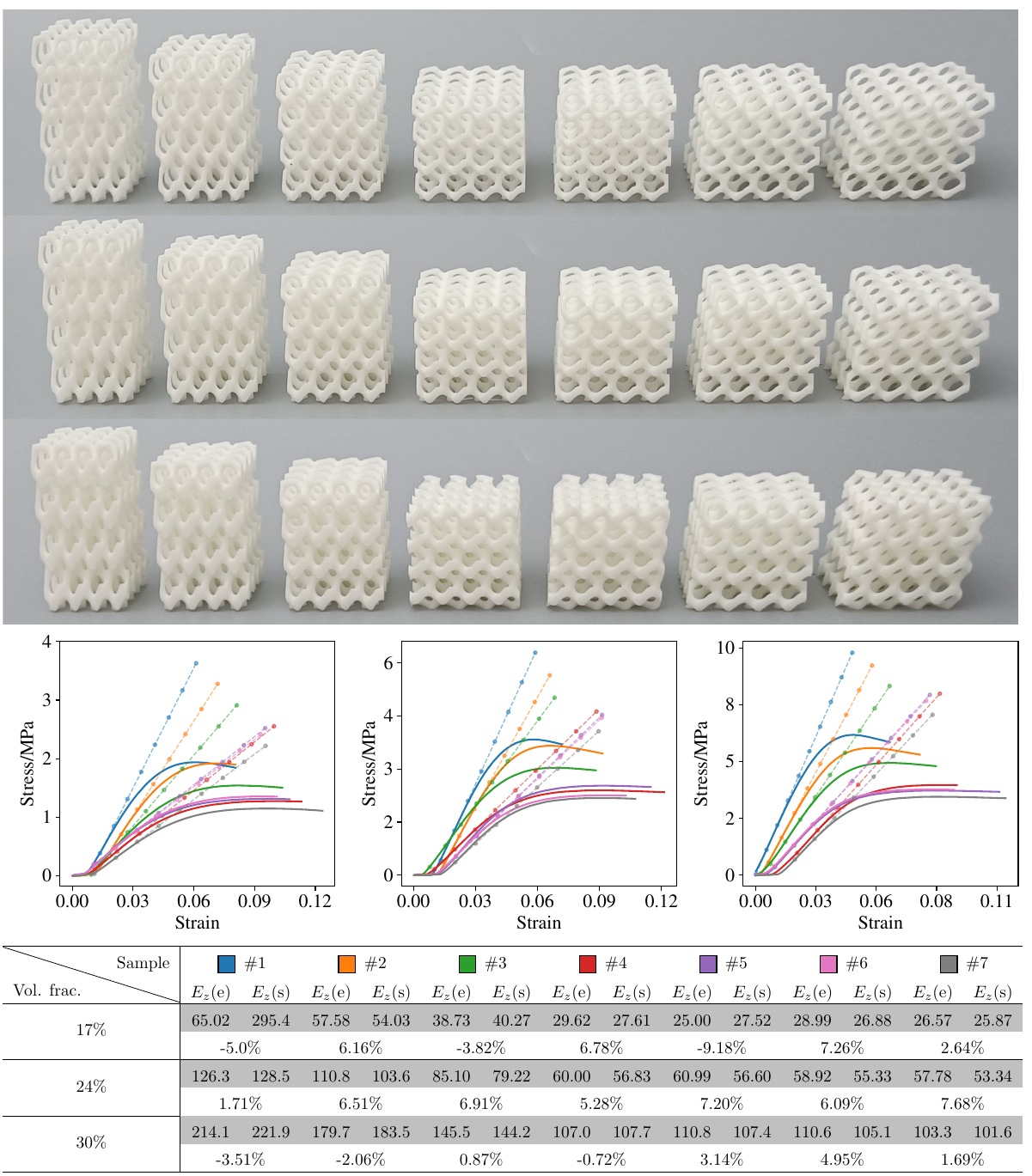}
    \caption{Physical experiments for different volume fractions. Top: test samples with type TPMS-TG while volume fraction in terms of 17\%, 24\%, 30\%, middle: strain-stress curves of z-direction (solid line) and their homogeneous prediction of linear part (dashed line) for different volume fractions; bottom: comparison of Young's modulus of z-directions w.r.t physical experiments $E_z(e)$ and PH-Net simulation $E_z(s)$ and their relative errors for different volume fractions.}
    \label{fig:diff_vf}
\end{figure*}

To validate the efficiency of PH-Net, we design a set of physical experiments w.r.t. different base materials, microstructure types and volume fractions.  PH-Net are well pre-trained to match different tasks. Meanwhile, each group of test samples as shown in Figure~\ref{fig:diff_material},~\ref{fig:diff_type} and~\ref{fig:diff_vf} include seven microstructures with the same microstructure types but different boundary shapes whose shape parameters are listed in Table~\ref{tab:sample_param}. All microstructure are repeated to $4\times4\times4$ blocks and printed 40mm in size.
We performed compression experiments in Z-direction on the electromechanical universal testing machine MTS E45.105, with the maximum rated force capacity 100kN.

The first experiment is to test PH-Net in terms of different base materials. As demonstrated in Figure~\ref{fig:diff_material}, we 3D printed microstructures using Thermoplastic Polyurethane (TPU), resin and metal (AlSi10Mg) materials, whose Young's Modulus are $24\mathrm{MPa}$, $750\mathrm{MPa}$ and $42\mathrm{GPa}$, respectively. For TPU material, the Poisson ratio is $0.45$, while others are both 0.25. The middle plots of Figure~\ref{fig:diff_material} illustrate the strain-stress relationship from the compression test, in which the solid curves are the experiment results, and the dashed lines are the corresponding simulation results.
Note that all microstructures except those made of metal reached their yield points because the metal microstructures exceed the measuring range of testing equipment before reaching their yield point. 
However, the available curves are sufficient for estimating the linear elasticity of metallic microstructures.
The tested and simulated Young's modulus $E_z(e)$ and $E_z(s)$ and their errors are listed in the bottom table of Figure~\ref{fig:diff_material}.
It can be seen that errors for all samples and materials are no more than 10\%, which is reasonable considering the fabrication and experiment errors. 
In contrast, the prediction error between PH-Net and numerical homogenization can be ignored in additive manufacturing.

We also design the test task for PH-Net in terms of different microstructure types and volume fractions. All microstructures are 3D printed with the same resin material whose Young's modulus is $[1200\mathrm{MPa}, 1400\mathrm{MPa}]$ in range and Poisson ratio is $0.25$. However, as different manufacturing batches may lead to different base material properties, we use test strips for each sample group to calibrate their base material properties. 
Figure~\ref{fig:diff_type} shows the testing results w.r.t Octet truss, TPMS-TG, and Kelvin truss with the same volume fraction 24\%. 
Figure~\ref{fig:diff_vf} demonstrates the testing results for TPMS-TG microstructures with different volume fractions.
Similar to the task on different base materials, the errors between physical and simulation results for both tasks are within $10\%$, which means our PH-Net is accurate enough.

\begin{table}
\centering
\scriptsize
\begin{tabular}{c|ccccccc} 
\toprule
\multicolumn{1}{l|}{\diagbox{Parameter}{Sample}} & \#1 & \#2 & \#3 & \#4 & \#5 & \#6 & \#7 \\ 
\hline
$\alpha_xy$ & $90^\circ$ & $90^\circ$ & $90^\circ$ & $90^\circ$ & $90^\circ$ & $90^\circ$ & $90^\circ$ \\
$\alpha_yz$ & $90^\circ$ & $90^\circ$ & $90^\circ$ & $90^\circ$ & $85^\circ$ & $80^\circ$ & $75^\circ$ \\
$\alpha_xz$ & $90^\circ$ & $90^\circ$ & $90^\circ$ & $90^\circ$ & $90^\circ$ & $90^\circ$ & $90^\circ$ \\
$l_x$ & 1 & 1 & 1 & 1 & 1 & 1 & 1 \\
$l_y$ & 1 & 1 & 1 & 1 & 1 & 1 & 1 \\
$l_z$ & 2 & 1.67 & 1.33 & 1 & 1 & 1 & 1 \\
\bottomrule
\end{tabular}
\caption{Shape parameters of test samples}
\label{tab:sample_param}
\end{table}

\section{Conclusion}

\begin{table}
\centering

\renewcommand{\arraystretch}{1.2}
\resizebox{\linewidth}{!}{%
\begin{tabular}{ccccccc} \toprule
\multirow{2}{*}{Method} & \multirow{2}{*}{On-the-fly} & \multicolumn{3}{c}{Generalization} & \multirow{2}{*}{\begin{tabular}[c]{@{}c@{}}Microscopic\\~properties\end{tabular}} & \multirow{2}{*}{Label-free} \\ \cline{3-5}
 &  & \begin{tabular}[c]{@{}c@{}}Microstructure \\type\end{tabular} & \begin{tabular}[c]{@{}c@{}}Base\\~material\end{tabular} & \begin{tabular}[c]{@{}c@{}}Boundary \\shape\end{tabular} &  &  \\ \hline
Numerical homogenization & $\times$ & $\checkmark$ & $\checkmark$ & $\checkmark$ & $\checkmark$ & - \\
Microstructure-to-material map & $\checkmark$ & $\times$ & $\times$ & $\times$ & $\times$ & $\times$ \\
Current ML/DL methods & $\checkmark$ & $\checkmark$ & $\times$ & $\times$ & $\times$ & $\times$ \\
PH-Net & $\checkmark$ & $\checkmark$ & $\checkmark$ & $\checkmark$ & $\checkmark$ & $\checkmark$ \\ \bottomrule
\end{tabular}
}
\caption{Comparison of numerical homogenization, current data-driven methods and PH-Net w.r.t. different respects.}
\label{tab:method_comparison}
\end{table}

In this paper, we propose PH-Net, a U-Net style convolutional neural network, to predict the homogenized material properties of parallelepiped microstructures.
Unlike existing DL methods, PH-Net predicts the microscope displacements under macroscope strains and periodic boundary conditions.
We construct a novel loss function based on MPE theory, with no need for ground-truth labels in dataset construction.
We also introduce a new dataset construction method for parallelepiped microstructures, encoding microstructure types, base materials and boundary shapes together. 
PH-Net can speed up the prediction of homogenized material properties hundreds of times. 
More importantly, PH-Net is label-free, superior to existing DL methods that require highly time-consuming ground-truth dataset construction.
PH-Net can evaluate many mechanical properties using displacements, e.g., worst-case stress distribution, yield strength, and shear strength, which are unavailable for other DL methods output elastic properties directly.
Besides, PH-Net takes the generalization of both microstructure types, their base materials and boundary shapes into account. Meanwhile, it is backwards compatible with existing DL methods as well.
A summary of PH-Net advantages compared to numerical homogenization method,  current explicit and implicit data-driven methods are list in Table~\ref{tab:method_comparison}.
PH-Net still has improvement space.
First, the shape parameters in the dataset are within $[75^{\circ},90^{\circ}]$ for angles and $[1,2]$ for scaling. We expect to expand the range of shape parameters and introduce more types of microstructures in the future.
Besides, the prediction error of PH-Net is much higher than the numerical homogenization method, and its training time is still rather long. 
We note that the computation of loss function still has improvement space, e.g., to reduce the illness of loss function, and the first-order Adam optimizer also has improvement space.
In the future, we plan to use numerical strategies, like building pre-conditioners to boost solving efficiency and implement higher-order optimizers, i.e., L-BFGS, to reduce iteration steps.

 \bibliographystyle{elsarticle-num} 
 \bibliography{ph-net}
\appendix
\section{Shape-material transformation for parallelepiped microstructures}
\noindent
\textbf{Proof}:
If and only if the boundary shape $\Omega$ belongs to parallelepiped, we have $J_\Omega=J_e$ for each element $e\in\Omega$.
Let $\mathcal{M}=JJ^T$ represent the transformation of second-order tensor for simplification, then $\hat{\varepsilon}=\mathcal{M}\varepsilon,\ \hat{\sigma}=\mathcal{M}\sigma$. For fourth-order material tensor, the shape-material transformation can be expressed as
\begin{gather}
    \hat{C}_{ijkl}=J^{-1}_{ip}J^{-1}_{jq}J^{-1}_{kr}J^{-1}_{ls}C_{pqrs}\iff \hat{C}=\mathcal{M}^{-T}C\mathcal{M}^{-1}\notag\\
    C^H_{ijkl}=J_{ip}J_{jq}J_{kr}J_{ls}\hat{C}^H_{pqrs}\iff C^H=\mathcal{M}^T\hat{C}^H\mathcal{M}
\end{gather}

Suppose $B_e$ and $\hat{B}_e$ are the shape matrices in term of parallelepiped $Q$ and cube $P$, the shape transformation is $\hat{B}_e=\mathcal{M}B_e$. With the same local displacement field $u$, the corresponding microscopic strains are $\hat{\varepsilon}(u)=\mathcal{M}\varepsilon(u)$.

First of all, substituting $\hat{\mathcal{C}}^b=\mathcal{M}^{-T}C^b\mathcal{M}^{-1}$ to the localization step of homogenization processing $\hat{\mathcal{H}}(\hat{C}^b,P)$, according to Equation~\ref{eq:discrete_localization}, we have
\begin{gather}
\sum_{e\in\Omega}\hat{B}_e^T\hat{C}^b\hat{B}_e\hat{u}_e\mathrm{d}\Omega=\sum_{e\in\Omega}\hat{B}_e^T\hat{C}^b\varepsilon^0\mathrm{d}\Omega\notag\\
\Updownarrow\notag\\
\sum_{e\in\Omega}\hat{B}_e^T\mathcal{M}^{-T}C^b\mathcal{M}^{-1}\hat{B}_e\hat{u}_e\mathrm{d}\Omega=\sum_{e\in\Omega}\hat{B}_e^T\mathcal{M}^{-T}C^b\mathcal{M}^{-1}\varepsilon^0\mathrm{d}\Omega\notag\\
\Updownarrow\notag\\
\sum_{e\in\Omega}B_e^TC^b\mathcal{M}^{-1}\hat{B}_e\hat{u}_e\mathrm{d}\Omega=\sum_{e\in\Omega}B_e^TC^b\mathcal{M}^{-1}\varepsilon^0\mathrm{d}\Omega\notag\\
\Updownarrow\notag\\
\sum_{e\in\Omega}B_e^TC^b\hat{B}_e\hat{u}_e\mathrm{d}\Omega=\sum_{e\in\Omega}B_e^TC^b\varepsilon^0\mathrm{d}\Omega.
\label{eqshape_material_tranformation_localization}
\end{gather}
At the same time, for homogeneous material $C^H=\mathcal{H}(C^b,P)$ defined on the parallelepiped $P$, its localization step can be expressed as
\begin{equation}
    \sum_{e\in\Omega}B_e^TC^b B_e u_e\mathrm{d}\Omega=\sum_{e\in\Omega}B_e^TC^b\varepsilon^0\mathrm{d}\Omega.
    \label{eqparallel_hexahedral_localization}
\end{equation}
Without loss of generality, Equation~\ref{eqshape_material_tranformation_localization} can be conducted to Equation~\ref{eqparallel_hexahedral_localization}, if $\hat{\varepsilon}(\hat{u}_e)=\hat{B}_e\hat{u}_e=B_e u_e=\varepsilon(u_e)$ for each element $e\in \Omega$. Thus the relationship between macroscopic strain field and local strain field in terms of $C^H=\mathcal{H}(C^b,Q)$ and $\hat{C}^H=\hat{\mathcal{H}}(\hat{C}^b,P)$ are $A_e=\hat{A}_e, \forall e\in\Omega$.

Secondly, the integration step of $C^H=\hat{H}(C^b,Q)$ can be conducted by
\begin{gather}
    C^H=\dfrac{1}{|\Omega|}\sum_{e\in\Omega}A^TC^bA\mathrm{d}\Omega\notag\\
    \Updownarrow\notag\\
    C^H=\dfrac{1}{|\Omega|}\sum_{e\in\Omega}\hat{A}^TC^b\hat{A}\mathrm{d}\Omega\notag\\
    \Updownarrow\notag\\
     C^H=\dfrac{1}{|\Omega|}\sum_{e\in\Omega}\hat{A}^T\mathcal{M}^T\hat{C}^b\mathcal{M}\hat{A}\mathrm{d}\Omega\notag\\
    \Updownarrow\notag\\
     C^H=\mathcal{M}^T\left[\dfrac{1}{|\Omega|}\sum_{e\in\Omega}\hat{A}^T\hat{C}^b\hat{A}\mathrm{d}\Omega\right]\mathcal{M}\notag\\
    \Updownarrow\notag\\
     C^H=\mathcal{M}^T\hat{C}^H\mathcal{M}.
\end{gather}
Done.





\end{document}